\documentclass[%
 reprint,
 amsmath,amssymb,
 aps,
nofootinbib,
]{revtex4-2}

\usepackage{graphicx}
\usepackage{dcolumn}
\usepackage{bm}
\usepackage{comment}
\usepackage[caption=false]{subfig}
\usepackage{xcolor}
\usepackage{float}

\usepackage{hyperref}
\hypersetup{
    pdftitle={A new determination of the (Z,A) dependence of coherent muon-to-electron conversion},
    pdfauthor={L\'eo Borrel, David G. Hitlin, Sophie Middleton},
    colorlinks=true,
    linkcolor=blue,
    citecolor=green,
    filecolor=blue,
    urlcolor=blue
}

\long\def\ignore#1{\relax}
\newcommand\Tstrut{\rule{0pt}{2.5ex}}  
\newcommand\Tstrutt{\rule{0pt}{3ex}} 
\usepackage{longtable}

\makeatletter

\renewcommand\paragraph{\@startsection{paragraph}{4}{\z@}%
                                     {-3.25ex\@plus -1ex \@minus -.2ex}%
                                     {1.5ex \@plus .2ex}%
                                     {\normalfont\normalsize\bfseries}}

\makeatother
 \let\paragraph\subparagraph \let\subparagraph\undefined

\begin{document}

\preprint{APS/123-QED}

\title{A New Determination of the \texorpdfstring{$(Z,A)$}{(Z,A)} Dependence of Coherent Muon-to-Electron Conversion}

\author{L\'eo Borrel}
\author{David G. Hitlin}
\author{Sophie Middleton}
\affiliation{
 California Institute of Technology\\
 Pasadena, CA 91125 USA
}
\date{\today}

\begin{abstract}

Should muon-to-electron conversion in the field of a nucleus be found in the current generation of experiments, the measurement of the atomic number dependence of the process will become an important experimental goal. We present a new treatment of the (Z,A) dependence of coherent muon-to-electron conversion in 236 isotopes. Our approach differs from previous treatments in several ways. Firstly, we include the effect of permanent quadrupole deformation on the charged lepton flavor violating matrix elements, using the method of Barrett moments. This method also enables the addition of muonic X-ray nuclear size and shape determinations of the charge distribution to the electron scattering results used previously. Secondly, we employ the deformed relativistic Hartree-Bogoliubov theory in continuum (DRHBc) to calculate neutron-related matrix elements for even-even nuclei. This takes into account the quadrupole deformation of the neutron distributions and the fact that neutrons are, in general, in different shell model orbits than protons. The calculated conversion rates differ from previous calculations, particularly in the region of large permanent quadrupole deformation. Finally, we propose an alternative normalization of the muon-to-electron conversion rate, which is related more closely to what a given experiment actually measures, and better separates lepton physics from nuclear physics effects.

\end{abstract}

\maketitle

\section{\label{sec:introduction}{Introduction}}

The most stringent experimental upper limit on charged lepton flavor violation (CLFV) via muon-to-electron conversion ($\mu \rightarrow e$) in the field of a nucleus  ($R_{\mu e} \leq 7 \times 10^{-13}$ at $90\%$  C.L.) was set by the SINDRUM-II experiment using a gold target \cite{SINDRUMII2006}. The Mu2e \cite{Mu2eTDR} and COMET \cite{COMETTDR} experiments, both nearing data-taking, will use an aluminum target. The Mu2e experiment at Fermilab aims to reach a single event sensitivity (SES) of $\sim 3 \times 10^{-17}$ for the conversion rate relative to muon capture. The COMET experiment at J-PARC, has a planned sensitivity of  $ 2 \times 10^{-15}$ for Phase I and a similar sensitivity to Mu2e for Phase II~\cite{Lee2018}. CLFV limits derived from the decays $\mu^+ \to e^+ \gamma$ and $\mu^+ \to e^+e^-e^+$, mainly sensitive to dipole coupling, are currently the most stringent. The new generation of deacy experiments, MEG-II for $\mu^+ \to e^+ \gamma$ \cite{MEGII} and Mu3e for $\mu^+ \to e^+e^-e^+$ \cite{Mu3e2020}, are underway at PSI.

Should the decay or conversion experiments find evidence for CLFV decays or  $\mu\! \rightarrow\! e$ conversion in aluminum, interest will turn to understanding the effective operators that mediate CLFV at low energies.  This can be done by studying the Dalitz plot distributions of polarized muon or tau decays to three leptons \cite{Okada1999,Dassinger2007,Matsuzaki2007} or by measurements of the conversion rate with other nuclear targets, as the ($Z,A$) dependence of the conversion rate is sensitive to the Lorentz structure \cite{Kitano2002,Cirigliano2009,Cirigliano2022,Heeck2022}. Future proposed experiments, such as Mu2e-II \cite{Mu2e-II2022} and a conversion search at the Advanced Muon Facility \cite{Aoki2022}, will provide opportunities to measure  $\mu \rightarrow e$ conversion in other target materials.

\par
The $\mu \rightarrow e$ conversion process manifestly involves the nuclear physics of the target elements. We discuss herein several aspects of extracting the New Physics from the nuclear physics. In particular, we present a new systematic study of the atomic number dependence of the coherent conversion rate. We utilize the method of Barrett moments \cite{Barrett1970} to add a substantial amount of muonic X-ray data to previous studies that mainly utilize electron scattering data on nuclear charge distributions. This allows us to take into account the fact that many nuclei have permanent quadrupole deformations, which affects the evaluation of matrix elements of New Physics operators. We also employ a new treatment of the neutron distribution and present a new approach to normalization of experimental and theoretical results.
\par
The study of the ($Z,A$) dependence of conversion is most cleanly done with a target consisting of a single stable isotope, as with $^{27}_{13}$Al, the target in both the Mu2e and COMET experiments.
Potential higher $Z$ targets such as Ti have many stable isotopes and are typically not readily available as separated isotopes in sufficient quantity to meet experimental conversion sensitivity requirements.  Therefore, in Sec.~\ref{sec:atomic_number_dependence} we discuss conversion rates for both targets composed of individual isotopes and targets with natural elemental abundance.
Section~\ref{sec:Normalization} addresses the question of the historical approach to the presentation of experimental limits on $\mu \rightarrow e$ conversion. The usual approach, in analogy to the concept of a decay branching fraction, is to normalize the conversion rate to the rate of muon capture on a particular nucleus, and therefore to present limits on this ratio, referred to as $R_{\mu e}$:
\begin{equation}
    R_{\mu e}(Z,A) = \frac{\Gamma(\mu^{-} + N(Z,A) \rightarrow e^{-} + N(Z,A))}{\Gamma(\mu^{-} + N(Z,A) \rightarrow \text{all captures})}.
    \label{eq:rmue}
\end{equation}

We introduce an alternative normalization, which hews more closely to what a given experiment actually measures and better separates the lepton physics and the nuclear physics, removing extraneous contributions to the $Z,A$ dependence of conversion. 

Section \ref{sec:target_optimization} is a discussion of the practicalities involved in choosing future targets.

\section{Review of existing literature}
\label{sec:review}

Should CLFV be observed in $\mu \to e$ conversion in the upcoming round of experiments, interest will turn to identifying the Lorentz structure of the CLFV coupling, which can be accessed by a sufficiently precise measurement of the atomic number dependence of the conversion rate \cite{Kitano2002,Cirigliano2009,Heeck2022,Davidson2018}.

The most widely cited treatment of the atomic number dependence of coherent $\mu \to e$ conversion is that of Kitano, {\it et al.}\cite{Kitano2002}, extended by Cirigliano {\it et al.}\cite{Cirigliano2009}, which includes 55 isotopes whose charge distribution, measured using elastic electron scattering, were compiled by De Vries {\it et al.}\cite{deVries1987}. These measurements are parameterized using a variety of models for the nuclear charge distribution: harmonic oscillator, two and three-parameter spherical Fermi, two and three-parameter Gaussian, as well as model-independent sum-of-Gaussians, and Fourier-Bessel expansion treatments where available. The spin-independent contribution to the coherent conversion rate has an $A^{2}$ enhancement and is in most models dominant.  Three types of effective operators, scalar, vector and dipole, contribute to coherent conversion. The rate of $\mu \to e$ conversion is calculated using the overlap integrals between the $\mu$ and $e$ Dirac wave functions and nucleon densities for these three cases. While some direct measurements of neutron distributions are explored in Refs.\cite{Kitano2002,Cirigliano2009}, for their main results the neutron and proton distributions are assumed to be identical, which means that the overlap integral related to the neutron distribution is equal to the overlap integral related to the proton distribution scaled by $N/Z$.

The recent calculations of Haxton {\it et al.}~\cite{Haxton2022}\cite{Haxton2024} improve these analyses for a selected number of nuclei by employing nuclear-level effective field theory to determine the matrix elements.
Effective field theory has also been exploited to refine the interpretation of experimental CLFV limits in terms of New Physics couplings \cite{Davidson2022}.

The Kitano {\it et al.} and Cirigliano {\it et al.} studies present the $Z$-dependence of the conversion rates for 55 isotopes, that have had their muon capture lifetime measured \cite{Measday2001}, in the form of a ratio: $ R_{\mu e}^{Z}/R_{\mu e}^{Al}$. The $Z$-dependence of this ratio has a great deal of structure, largely due to the influence of ``magic numbers" associated with closed shells in the nuclear shell model on the nuclear size and shape. The influence of shell structure on the nuclear size, and thus the evaluation of the New Physics matrix elements for the largely coherent conversion process is unavoidable. The additional structure in the $Z$ dependence, introduced by the division by the muon capture rate, a non-coherent Standard Model process, can, however, be avoided. This formulation is traceable in part to the original conversion normalization proposal of Weinberg and Feinberg \cite{Weinberg1959}. The question of normalization of the conversion rate will be discussed in Sec. \ref{sec:Normalization}.

\subsection{Treatment of the neutron distribution}

Kitano {\it et al.} \cite{Kitano2002} used measured neutron distributions when possible. These measurements, derived from charged pion, proton or alpha particle scattering, were available for only 16 nuclei.
Their primary method of including neutrons in determining the $(Z,A)$ dependence of conversion was therefore to scale the overlap integral related to the proton distribution by a factor of $N/Z$. Particularly in heavy nuclei, protons and neutrons populate levels with different quantum numbers and are thus in different shells. This can produce rather different shapes and sizes for protons and neutrons, especially for nuclei at the edge of the valley of stability. In particular quadrupole deformations, as measured by  $B$(E2) (the electric quadrupole transition probabilities) determined by Coulomb excitation and by muonic X-ray hyperfine structure, show a different dependence for protons and neutrons \cite{Stelson1965}. We have therefore resorted to a model calculation to account for these details. 
To improve on simple $N/Z$ scaling, we account for the fact that neutron distributions are typically larger than proton distributions by using a comprehensive set of calculations employing the deformed relativistic Hartree-Bogoliubov theory in continuum (DRHBc) \cite{Zhou2009, Li2012} to estimate the size and shape of the neutron distributions. This approach more accurately accounts for the fact that in most nuclei, particularly those in the region of large permanent quadrupole deformation ($Z = 60-80$), neutrons are in different shells than protons.
 Zhang {\it et al.}~\cite{Zhang2022} have used the DRHBc theory~\cite{Zhou2009, Li2012} for even-even nuclei ranging from $Z=8$ to $Z=120$ to determine individual neutron and proton distributions, accounting for quadrupole deformation. This work has recently been extended to other even-nuclei~\cite{Guo2024}, but it has not been included in the results we are showing here.

\subsection{The effects of quadrupole deformation}

Many nuclei, especially those with protons or neutrons far from a closed shell, are non-spherical, having permanent quadrupole or higher multipole shapes. We address the effect of these deformations on the calculation of $\mu \rightarrow e$ conversion rates. The conversion $\mu \rightarrow e$ occurs primarily from the 1$S$ state, so conversion rates can be calculated using only the radial Dirac equation with an equivalent spherically symmetric charge (or neutron) distribution. However, characterizing the average over non-spherical distributions with a derived $rms$ radius does not account for the effective smearing at the outer edge of a nucleus due, for example, to a quadrupole deformation.

The effective nuclear charge or neutron distribution skin thickness enters into the evaluation of the CLFV matrix elements. The relative importance of the skin thickness in determining the matrix element depends on the atomic number ({\it c.f.} $N/Z$) and the quadrupole deformation.
We calculate the CLFV matrix elements using a Dirac equation with spherical symmetry, the angular integration accounting for the effect of the quadrupole deformation, whether oblate or prolate. This produces an effect largely concentrated at the nuclear surface, increasing the effective skin thickness of the charge distribution. The $N/Z$ ratio increases with atomic number; the excess neutrons producing a ``neutron skin'', a larger effective skin thickness for the neutrons, as well as effects on the charge distribution, {\it e.g.} isotope shifts in electronic or muonic atom spectra. These effects are further modulated by both proton and neutron shell structure, particularly around magic number closed shells.

The case of $^{27}$Al is of particular interest, as it is the target for the Mu2e and COMET experiments.  $^{27}$Al has a large quadrupole moment (146.6 mb) ~\cite{Pyykko2008}. Most shell model calculations conclude that this nucleus has a prolate shape, with $\beta = 0.39$, although Dehnhard~\cite{Dehnhard1972} argues for an oblate ground state deformation. The sign of the quadrupole deformation does not enter into the calculation of the relevant integrals in the solution of the Dirac equation.

\subsection{The isotopes of neodymium}

\begin{figure*}[!ht]
    \centering
    \includegraphics[width=\textwidth]{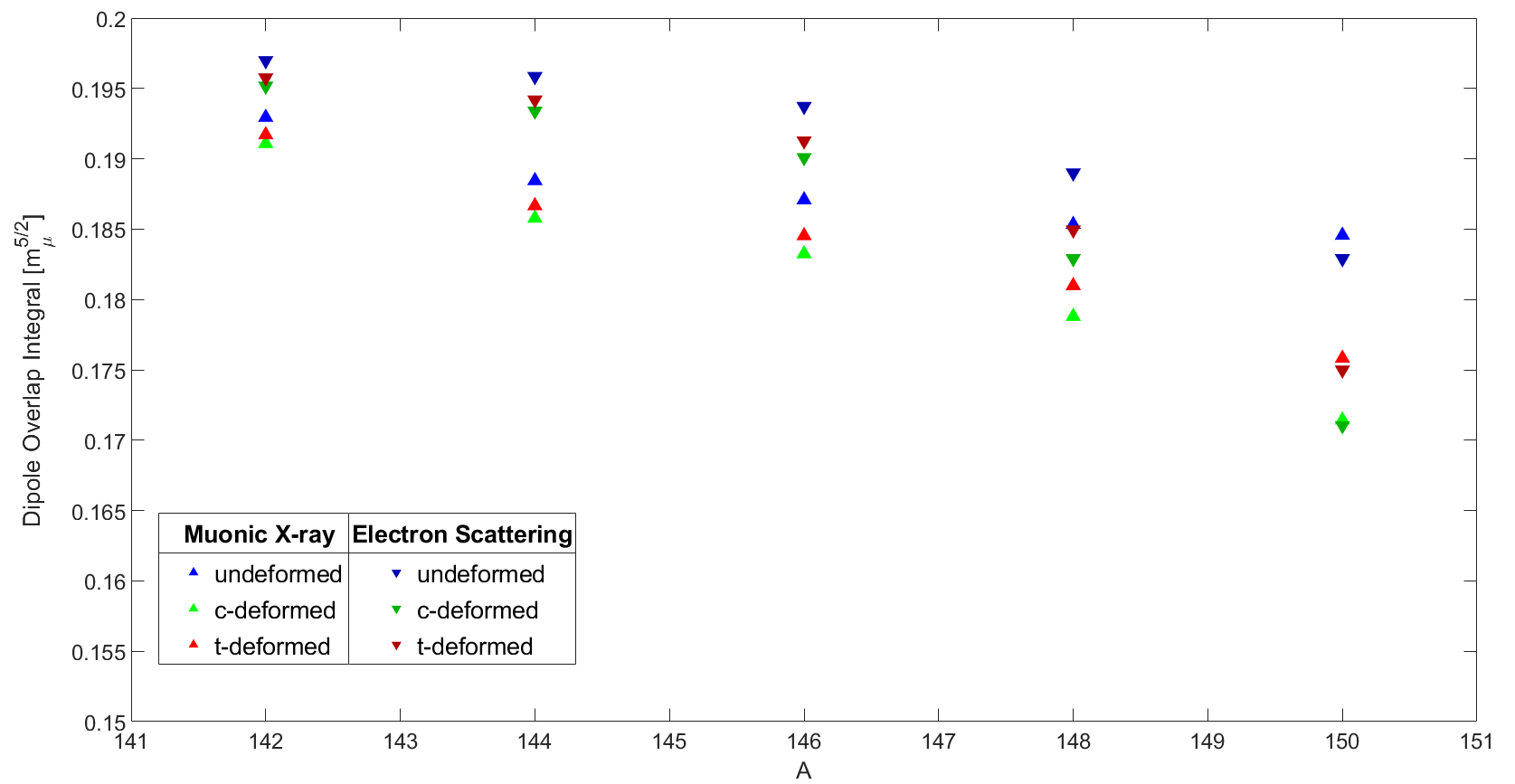}
    \caption{The dipole overlap integral for five even-even isotopes of Nd as determined using muonic X-ray data~\cite{Macagno1970} and electron scattering data~\cite{Heisenberg1971} and three treatments of the charge distribution (detailed in Sec. \ref{sec:deformation}).}
    \label{fig:Nd_study}
\end{figure*}

Neodymium has five significant stable isotopes whose charge distributions have been measured in both
electron scattering \cite{Heisenberg1971} and muonic X-ray spectra \cite{Macagno1970}. This provides an opportunity to explore some of the subtleties encountered in combining these different determinations of the nuclear charge distribution. There are measurements on five even-even neodymium isotopes: $A$=142, 144, 146, 148, and 150, which range from spherical to deformed. This allows us to explore the interplay of the effect of a quadrupole deformation on the effective skin thickness, as well as the differences between the two techniques, as shown in Fig.~\ref{fig:Nd_study}.

\begin{figure*}[!ht]
    \centering
    \includegraphics[width=\textwidth]{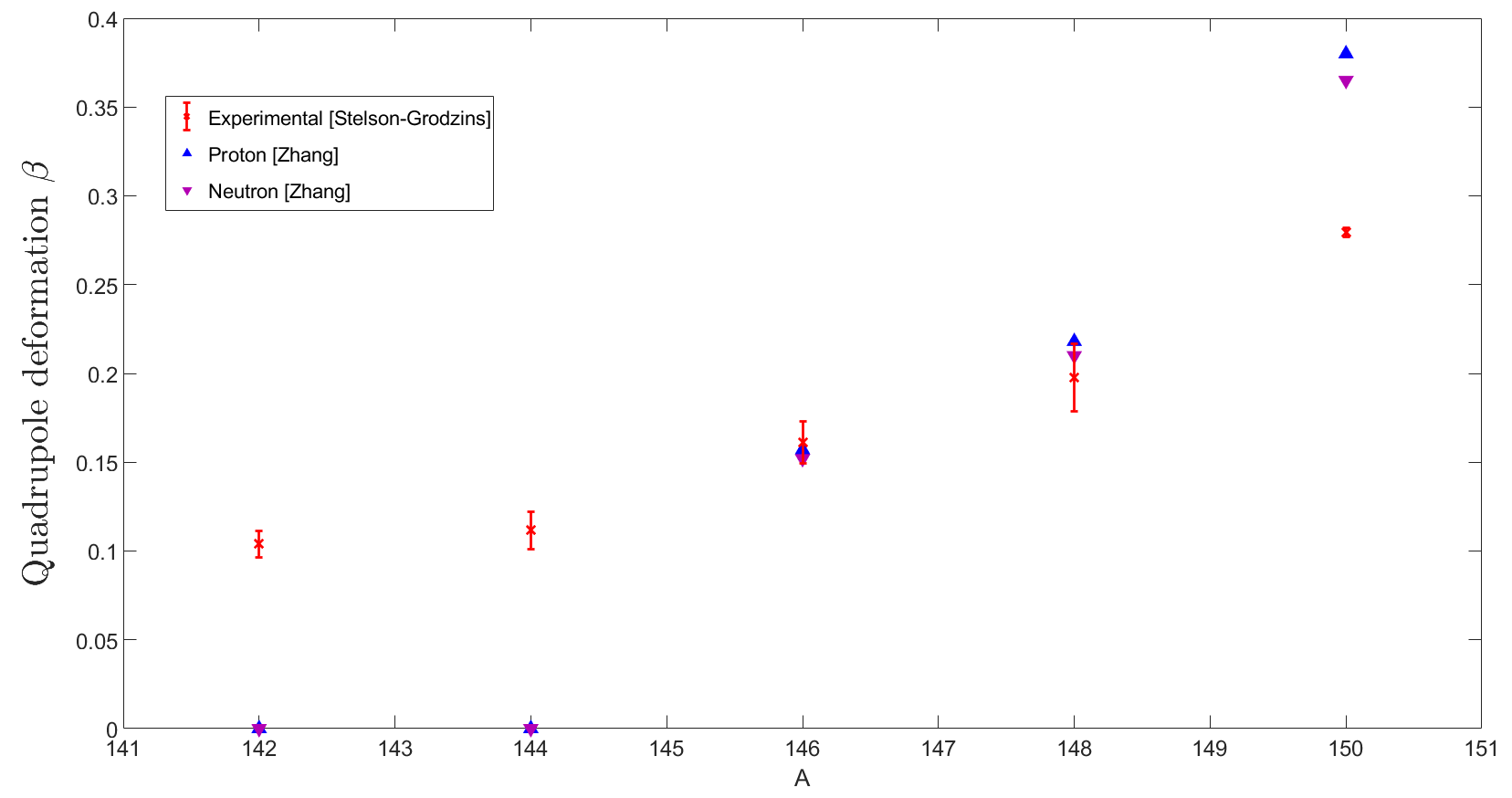}
    \caption{Quadrupole deformation of five even-even isotopes of Nd determined from $B$(E2) measurements in Coulomb excitation~\cite{Stelson1965} (with experimental uncertainty), and calculated for the charge and neutron distributions in the model of Zhang {\it et al.}~\cite{Zhang2022}.} 
    \label{fig:Nd_study_b}
\end{figure*}

The values of the quadrupole deformation $\beta$, determined largely from $B$(E2) measurements \cite{Stelson1965}, show a steady increase with neutron number, culminating at $N = 90$ (shown in Fig. \ref{fig:Nd_study_b}). The value of $\beta$ in the three-parameter fits to the charge distribution is held fixed for both the muonic X-ray and electron scattering cases.

The calculations of Zhang {\it et al.} using the DRHBc theory show that the {\it rms} charge radius of these Nd isotopes grows as $A^{1/3}$, while the neutron radius grows quicker, as the neutrons are in a higher shell and further from a shell closure. This is our primary motivation for using the Zhang {\it et al.} calculations to evaluate the $S(n)$ and $V(n)$ overlap integrals, rather than scaling the neutron distributions by ${N/Z}$ as has been the practice in \cite{Kitano2002}\cite{Cirigliano2009}. Related calculations by Zhang {\it et al.}\cite{Pan2022} also reproduce the behavior of the quadrupole deformation in the Nd isotopes as a function of neutron number.

\subsection{Goal of this study}

Our new determination of the atomic number dependence of the coherent conversion rate employs the method of Barrett moments \cite{Barrett1970} to add the many muonic X-ray measurements of nuclear charge distributions to the electron scattering data used in previous studies. The muonic X-ray data encompass measurements of many strongly-deformed nuclei, using the resolved hyperfine structure in the X-ray transitions to measure the nuclear quadrupole moment shape, typically parameterized by a three-parameter Fermi distribution. 

Our treatment of the $A$ and $Z$ dependence of conversion differs from previous treatments in several ways:
\begin{itemize}
    \item While previous studies have mainly used electron-scattering-based nuclear size determinations, we include the many measurements of nuclear charge distributions determined using muonic X-rays. This substantially enlarges the sample size, particularly in the regime above $Z = 60$, where many nuclei have substantial quadrupole deformations. We combine the elastic electron scattering and muonic X-ray data using the method of Barrett moments, and devise a procedure to incorporate the effect of permanent $Y_{20}$ deformations on the effective nuclear skin thickness. By this approach, we can include the conversion rates for a total of 236 isotopes.
    
    \item Rather than using a neutron distribution equal to the proton charge distributions, the primary method used in previous treatments, we use the deformed relativistic Hartree-Bogoliubov theory in continuum (DRHBc) \cite{Zhang2022} for the neutron distributions and compare with other approaches. This model, bench-marked against a wide variety of experimental determinations for even-even nuclei, allows us to include a wider variety of nuclei and to explore isotopic effects on the conversion rate. 

    \item We present the $\mu \rightarrow e$ conversion rates for the coherent New Physics process normalized to the total muon lifetime, rather than to the muon capture rate.
    
\end{itemize}

\section{Atomic number dependence of \break \texorpdfstring{$\mu \rightarrow e$}{muon-to-electron} conversion}
\label{sec:atomic_number_dependence}

\subsection{Theoretical background}
\label{sec:theory}

In the following study, we concentrate on the dominant spin-independent coherent conversion process in which the final and initial state of the nucleus is identical and the coherent conversion rate is enhanced by $A^{2}$.

The lepton flavor-violating coherent conversion rate of a muon into an electron is given by Kitano \textit{et al.} \cite{Kitano2002} as:

\vspace{-30pt}
\begin{widetext}
    \begin{equation}
        \begin{gathered}
            \Gamma_{conv} = 
            2 G_F^2 \left| A^*_R D + \tilde{g}^{(p)}_{LS} S^{(p)} + \tilde{g}^{(n)}_{LS} S^{(n)} + \tilde{g}^{(p)}_{LV} V^{(p)} + \tilde{g}^{(n)}_{LV} V^{(n)} \right|^2 \\
            \phantom{XXX} + 2 G_F^2 \left| A^*_L D + \tilde{g}^{(p)}_{RS} S^{(p)} + \tilde{g}^{(n)}_{RS} S^{(n)} + \tilde{g}^{(p)}_{RV} V^{(p)} + \tilde{g}^{(n)}_{RV} V^{(n)} \right|^2
        \end{gathered}
    \label{eq:rate}
    \end{equation}
\end{widetext}

\newpage
\noindent where $G_F$ is the Fermi constant, and $A^*_{R,L}$ and $\tilde{g}$ are dimensionless coupling constants that describe the strength of each component of the Lagrangian.  $D,S$ and $V$ are the overlap integrals for dipole, scalar, and vector interactions respectively. The $L$ and $R$ subscripts indicate left-handed and right-handed components and the ($n$) and ($p$) superscripts denote the neutron and proton terms respectively.

The overlap integrals are defined as:

\begin{align*}
    D &= \frac{4}{\sqrt{2}} m_\mu \int_0^\infty \left( - E(r) \right) \left( g^-_e f^-_\mu + f^-_e g^-_\mu \right) r^2 dr \\
    S^{(p)} &= \frac{1}{2 \sqrt{2}} \int_0^\infty Z \rho^{(p)}(r) \left( g^-_e g^-_\mu - f^-_e f^-_\mu \right) r^2 dr \\
    S^{(n)} &= \frac{1}{2 \sqrt{2}} \int_0^\infty (A - Z) \rho^{(n)}(r) \left( g^-_e g^-_\mu - f^-_e f^-_\mu \right) r^2 dr \\
    V^{(p)} &= \frac{1}{2 \sqrt{2}} \int_0^\infty Z \rho^{(p)}(r) \left( g^-_e g^-_\mu + f^-_e f^-_\mu \right) r^2 dr \\
    V^{(n)} &= \frac{1}{2 \sqrt{2}} \int_0^\infty (A - Z) \rho^{(n)}(r) \left( g^-_e g^-_\mu + f^-_e f^-_\mu \right) r^2 dr \\
\end{align*}

\noindent where $f$ and $g$ are the lower and upper components of the radial solution of the Dirac equation describing a muon orbiting a nucleus and converting to an electron. The functions $f_\mu$ and ($f_e$) are the muon (electron) wavefunction~\cite{Rose1951}. The terms $\rho^{(n)}$ and $\rho^{(p)}$ describe the neutron and proton densities in the nuclei respectively. The electric field, $E(r)$, is derived using Gauss' Law:

\begin{equation}
   E(r) = \frac{Ze}{r^2} \int_0^r r'^2\rho^{(p)}(r') dr'.
\end{equation}

\begin{figure}[!ht]
    \centering
    \subfloat{\includegraphics[width=0.5\textwidth]{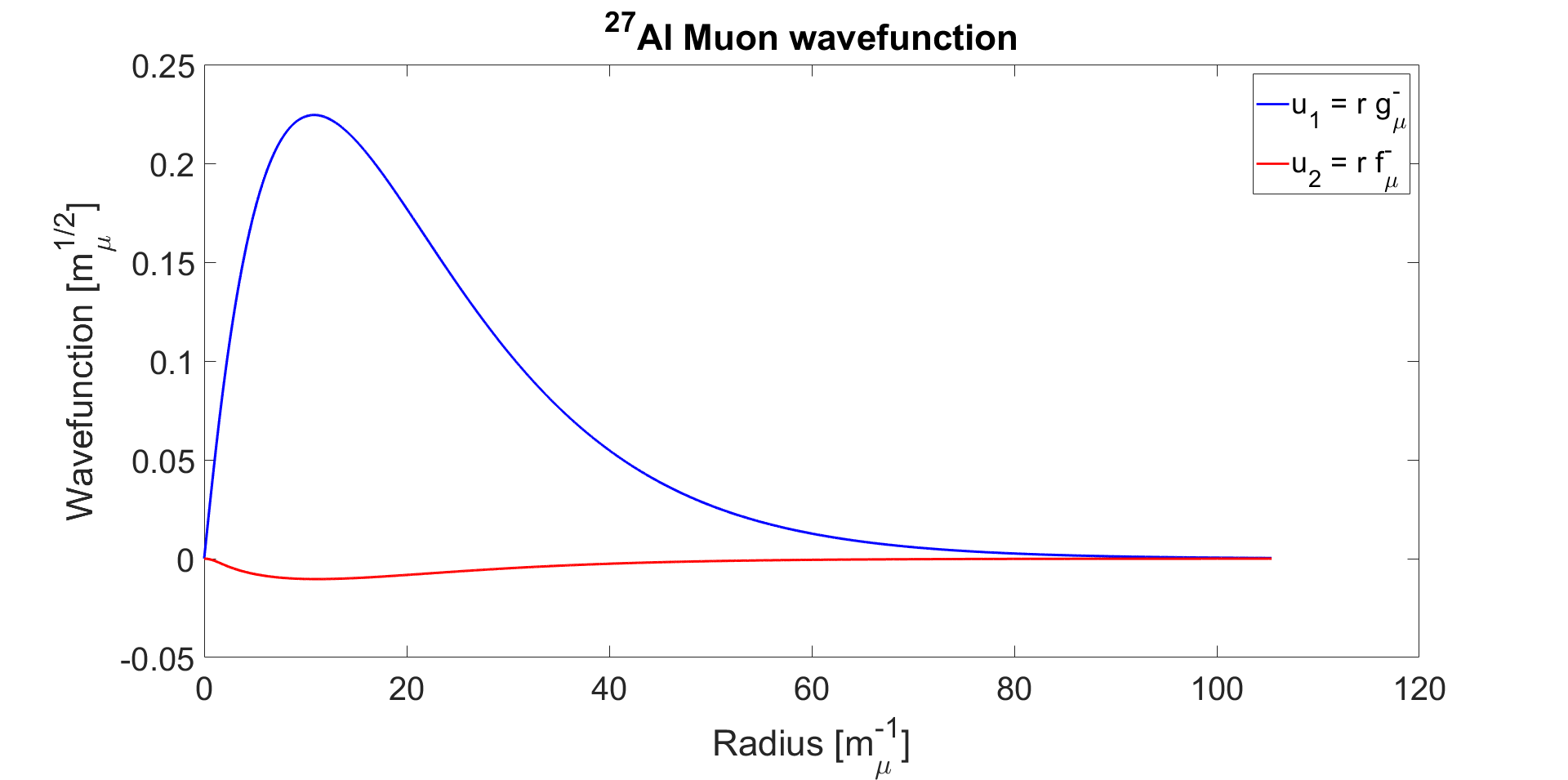}}\hfill
    \subfloat{\includegraphics[width=0.5\textwidth]{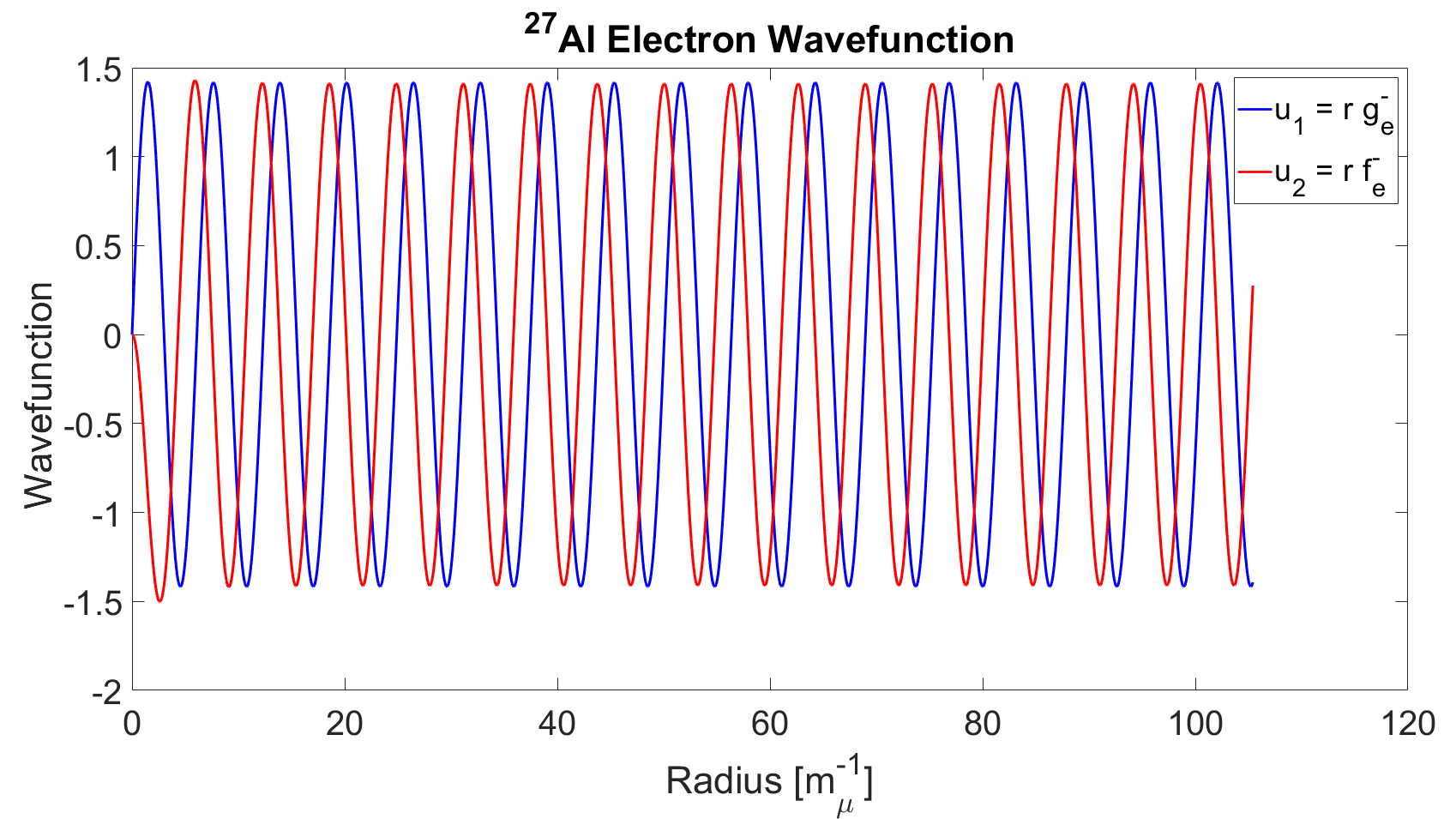}}
    \caption{The two components of the wavefunctions of the muon (top) and the electron (bottom) in the vicinity of the Al nucleus.}
    \label{fig:wavefunction}
\end{figure}

The muon and electron wavefunctions in the vicinity of the $^{27}$Al nucleus are shown in Fig. \ref{fig:wavefunction}. The conversion occurs overwhelmingly from the $1S$ state; we therefore integrate over polar and azimuthal angles to account for the effect of quadrupole deformations.

\subsection{Methodology}

\subsubsection{Combining electron scattering and muonic X-ray measurements}
\label{sec:method_combine}

To combine measurements of nuclear charge distribution obtained from electron scattering and muonic X-rays we use the method developed by Barrett \cite{Barrett1970}.
Electron scattering experiments derive information on the nuclear charge distribution from the energy and angular distribution of the scattered electron, mapping the momentum transfer of the electron to the Fourier transform of the charge distribution \cite{deVries1987}. These treatments typically keep only the first term in the $q^2$ expansion.

Muonic X-ray experiments use the energy of X-rays emitted in the atomic cascade of muonic atoms to determine the size and shape of the nucleus. Due to the strong overlap of the muon wavefunction with the nucleus, the energy of the $3D-2P$ and $2P-1S$ transitions is highly sensitive to nuclear size parameters. Analysis of the hyperfine structure allows for the (model-dependent) determination of permanent quadrupole charge distributions. This method typically provides a more precise measurement of the quadrupole deformation of nuclei than electron scattering experiments.

While for light nuclei the $2P-1S$ transition energy of a muonic atom can be matched to a single mean square radius $\left\langle r^2\right\rangle$, this is not true for heavier nuclei, where nuclear distributions with the same RMS radius can generate quite different transition energies.
Barrett \cite{Barrett1970} introduced a different moment, shared by all nuclear distributions, yielding the same transition energy. The Barrett moment is defined as:

\begin{equation}
    \left\langle r^k e^{-\alpha r}\right\rangle = \frac{4 \pi}{Z e} \int_0^\infty \rho(r) r^k e^{-\alpha r} r^2 dr,
    \label{eq:Barrett}
\end{equation}

\noindent where $k$ and $\alpha$, which are $Z$-dependent, are given in \cite{Barrett1970}.

We define the Barrett radius $R_{k \alpha}$ from the implicit equation \cite{Fricke1995}:

\begin{equation}
    \frac{3}{(R_{k \alpha})^3} \int_0^{R_{k \alpha}} \rho(r) r^k e^{-\alpha r} r^2 dr = \left \langle r^k e^{-\alpha r} \right \rangle.
\end{equation}

In both electron scattering and muonic X-ray analyses, the nuclear charge distribution has often been described using a 2-parameter Fermi distribution (2pF):

\begin{equation}
    \rho(r) = \frac{\rho_0}{1 + \exp \left[ \frac{r - c}{a} \right]}\ ,
    \label{eq:2pF}
\end{equation}

\noindent where $c$ is the radius at the half maximum and $t = 4 \, a \, \ln(3)$ is the skin thickness, defined as the region in which the charge density falls from 90\% to 10\% of the central value. This model is central to our Barrett moment-centered treatment.

We can see an example of the computation of the Barrett moment from two different datasets for neodymium in Table \ref{table:Nd}. The Barrett moments derived from the two types of experiments are seen to be in good agreement, supporting our approach to combining the electron scattering and muonic X-ray of measurements in determining the $Z$ dependence of the conversion rate.

\subsubsection{Quadrupole deformation}
\label{sec:deformation}

In order to compute the overlap integral, we need to solve the Dirac equation to obtain the muon and electron wavefunctions (as described in Sec. \ref{sec:theory}).
We used an existing code \cite{Bartolotta2017} to solve the Dirac equation for a muonic atom in the potential of a spherically-symmetric 2-parameter Fermi distribution.
For approximately spherical nuclei ($Z < 60$ and $Z > 80$), we can directly apply this solver, whereas for deformed nuclei we first need to get an equivalent spherically-symmetric distribution.

The method used to include the effect of nuclear quadrupole deformation on the matrix elements is as follows:

\begin{itemize}
    \item employ a deformed 3-parameter Fermi distribution\footnotemark[1]:
    \begin{equation}
    \label{eq:2pf_deformed}
        \rho(r, \theta) = \frac{\rho_0}{1 + \exp \left[ \frac{r - c (1 + \beta Y_{20}(\theta))}{a} \right]}
    \end{equation}
    \noindent where we take parameters $c$, $t = 4 \, a \, \ln(3)$ and $\beta$ from available literature detailed in Sec. \ref{sec:dataset}.
    \item We convert this to a spherically-symmetric 2-parameter Fermi distribution by removing the quadrupole deformation $\beta$ and adjusting the skin thickness $t$ to provide a constant Barrett moment (defined in Eq. \ref{eq:Barrett}).
    This equivalent 2-parameter Fermi distribution is then used to solve the Dirac equation.
\end{itemize}

\footnotetext[1]{This is not to be confused with the spherical three-parameter Fermi distribution often used in electrons scattering from heavy nuclei.}

Using the $t$ parameter to match the Barrett moment effectively smears out the nuclear surface, accounting for the integration of the $Y_{20}$ deformation over angles.

One could modify the $c$ parameter rather than the $t$ parameter to match the Barrett moment. This produces a small difference in the overlap integrals (see example for Nd in Fig. \ref{fig:Nd_study}).

\subsubsection{Dataset}
\label{sec:dataset}

Our calculations employ a compilation of data from different sources, which we list in the footnotes of the tables in Appendix \ref{sec:long_table}.
When there are different sources for data on an isotope, we use the data from the compilation listing the largest number of isotopes.

\begin{itemize}
    \item Nuclear charge distribution parameters:\\
        The majority of $c$ and $t$ values we employ come from the Fricke {\it et al.} \cite{Fricke1995} (Table III.A and III.C) compilation of muonic X-ray and electron scattering experimental data.
        For several specific isotopes that were not included in \cite{Fricke1995}, muonic X-ray measurements were added \cite{Landolt-Bornstein2004, Dewit1966, Hoehn1981, Zumbro1986, Zumbro1984}.
        When muonic X-ray measurements do not exist, we use the compilation of electron scattering experimental data from de Vries {\it et al.} \cite{deVries1987}.
        When the only information available is the {\it rms} charge radius\footnotemark[2], we use data from Angeli {\it et al.} \cite{Angeli2013}.
        \par
        For quadrupole deformation in the deformed region ($60 \leq Z \leq 80$), we use the Zhang {\it et al.} \cite{Zhang2022} model that calculates separate quadrupole deformations only for even-even proton and neutron distributions.
        For the the other deformed nuclei, we use the model of  M\"oller {\it et al.} \cite{Moller2016} for deformed isotopes. This employs a Finite-Range Drop Model (FRDM) to compute ground-state quadrupole deformations, as well as many other nuclear properties.

    \item Barrett moment parameters:\\
        There are two extensive compilations of the Barrett parameters $k$ and $\alpha$: Barrett {\it et al.}~\cite{Barrett1970} and Fricke {\it et al.}~\cite{Fricke1995}.
        The latter is fitted to the experimental muonic X-ray measurements that we also use, but it is not available for every isotope, whereas the former is theoretical but can be applied to every elements.
        As we only compute Barrett moments in deformed nuclei ($60 \leq Z \leq 80$), we decided to use the incomplete Fricke compilation.
        For isotopes that are not available, we chose to interpolate using neighboring elements.
        We decided not to use the Barrett compilation as there is a significant difference of up to $15 \%$ with the Fricke compilation.
        It is worth noting that even with this significant discrepancy between the two datasets, the difference in the computed overlap integrals is smaller than $0.25 \%$.
        
    \item Neutron distribution:\\
        We use the Zhang {\it et al.} \cite{Zhang2022} DRHBc model to separately evaluate the proton and the neutron nuclear distributions. This model treats only even-even nuclei, avoiding the additional single particle effects on both charge and neutron distributions seen in even-odd and odd-even nuclei.
        It provides different quadrupole deformation for the neutron and proton distributions.
\end{itemize}

\footnotetext[2]{If the available literature does not provide a 2-parameter Fermi distribution - some electron scattering measurements use a Fourier-Bessel expansion -, we use a least square method to fit the distribution to a 2pF assuming a fixed value of t = 2.3 fm}

\subsubsection{Uncertainties}
\label{sec:uncertainties}

We consider two sources of systematic uncertainties: the treatment of the  quadrupole deformation $\beta$ and the radius at half maximum $c$ in the Fermi distribution.

We take the quadrupole deformation error from Stelson and Grodzins~\cite{Stelson1965}, where uncertainty on $B$(E2) is converted into uncertainty on $\beta$ assuming a uniform ellipsoidal charge distribution.
The uncertainty on $c$ is taken directly from Fricke {\it et al.} \cite{Fricke1995}. 

Across all isotopes, the relative uncertainty on $c$ does not exceed $2 \%$, while the relative uncertainty on $\beta$ can be up to $20 \%$. We therefore propagated only the uncertainty in $\beta$ to the Barrett moment and ignored the $c$ uncertainty.

The resulting relative uncertainty on the Barrett moment on all the available elements never exceeded $1 \%$; we show the values for the neodymium isotopes in Table \ref{table:Nd} and Fig.~\ref{fig:Nd_study_b} as an example.

It should be noted that different sources for nuclear distribution parameters can have significant impact on the resulting overlap integral values.
We have tried to keep most of our data coming from a single compilation~\cite{Fricke1995} to limit this variation, but we had to add additional sources for missing data (visible in the footnotes of the tables in Appendix \ref{sec:long_table}).

\subsection{Results}

Appendix \ref{sec:long_table} contains all the results and parameters used in this study.
It is divided in two tables, Table \ref{table:undeformed} shows the results for nuclei that we approximate as spherically symmetric, and Table \ref{table:deformed} for the deformed nuclei ($60 \leq Z \leq 80$).
The components of the 2-parameters Fermi distributions of each isotope is provided, as well as the values of the Barrett parameters for deformed nuclei.

\begin{figure*}[!ht]
    \centering
    \includegraphics[width=1.0\textwidth]{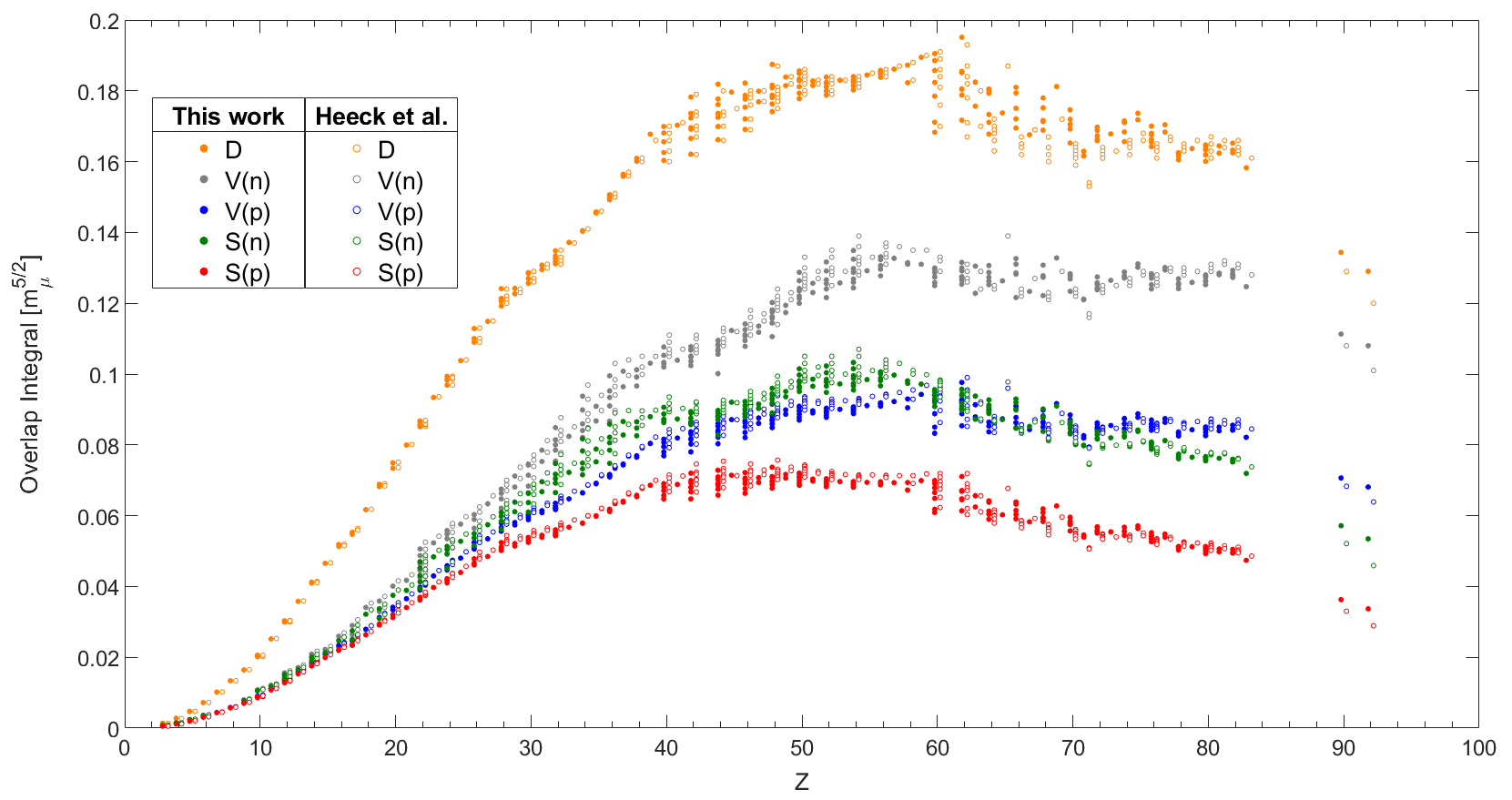}
    \caption{The overlap integrals for dipole, vector and scalar couplings as a function of atomic number for isotopes having a natural abundance $>$ 1\%, for the compilation of Heeck \textit{et al.}\cite{Heeck2022} and this work. For clarity, we shift the result of  Heeck \textit{et al.} by $+\frac{1}{2}$ unit on the x-axis.}
    \label{fig:OI_FM_all}
\end{figure*}

The resulting overlap integrals calculated on the 236 stable isotopes with natural abundance above $1 \%$ are displayed in Fig. \ref{fig:OI_FM_all}.
The plot compares the results of Heeck \textit{et al.} \cite{Heeck2022}; our new results are in general agreement.
The main difference appears in the deformed nuclei region ($60 \leq Z \leq 80$), where our values are higher due to our explicit treatment of the effect of the deformation on the matrix elements.
This results in a smoother behavior of the overlap integral value after $Z = 60$, furthering our attempts to isolate the New Physics CLFV interaction from nuclear effects.

There are a couple of outliers, for example $_{65}^{159}$Tb and $_{67}^{165}$Ho, that display smaller value compared to previous work, with significant difference in the case of $_{65}^{159}$Tb.
The discrepancy can be attributed to the origin of the nuclear distribution data, as these nuclei are not part of the main compilation of nuclear data used~\cite{Fricke1995}.
For $_{65}^{159}$Tb, choosing experimental data from deWit {\it et al.}~\cite{Dewit1966} compared to extrapolating 2pF parameters from the rms radius given in Angeli {\it et al.}~\cite{Angeli2013} accounts for the difference.

\begin{figure*}[!ht]
    \centering
    \includegraphics[width=1.0\textwidth]{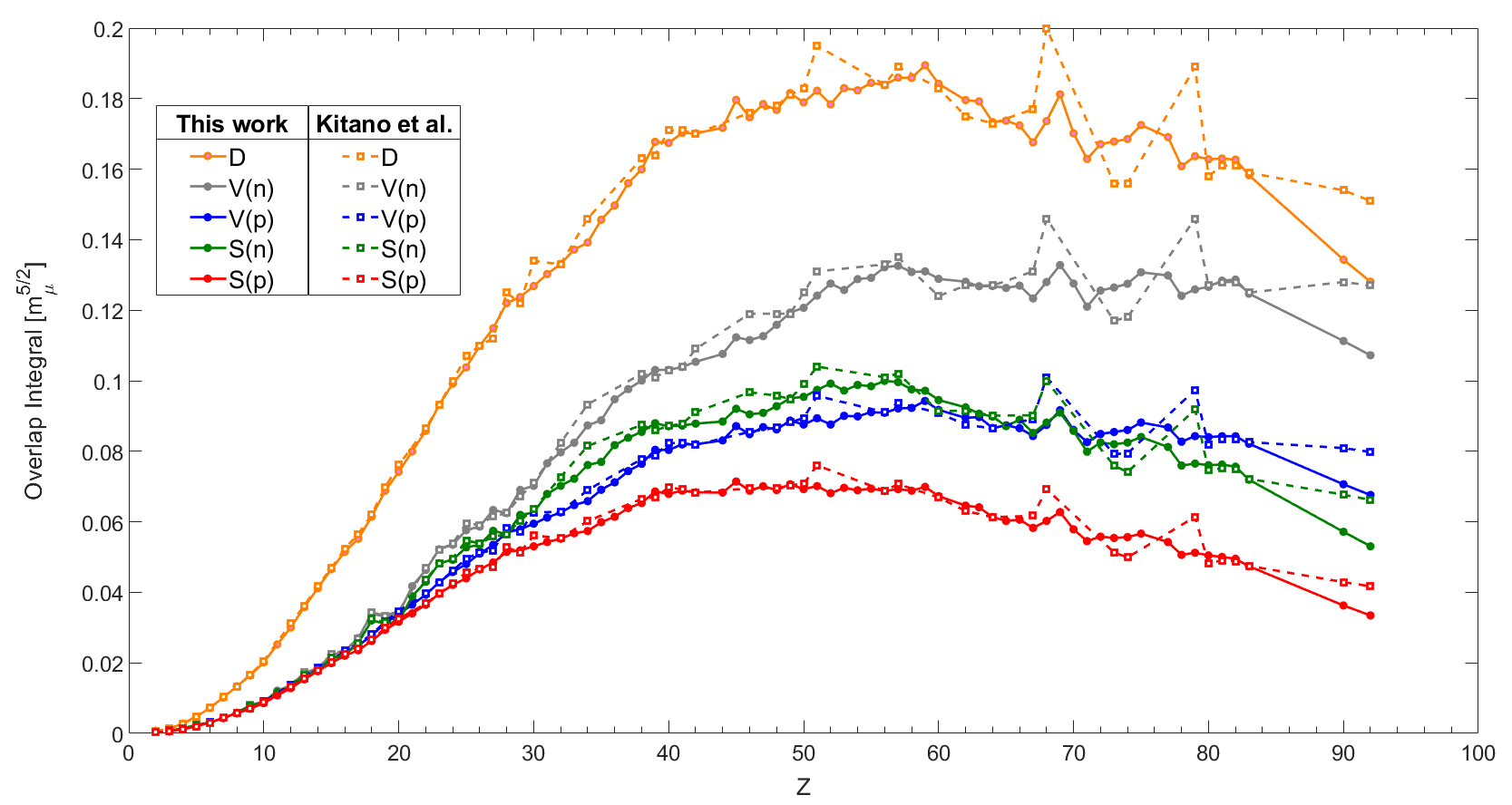}
    \caption{The overlap integrals for dipole, vector and scalar couplings as a function of atomic number for each element, weighted by natural abundance, for the compilation of Kitano {\it et al.}\cite{Kitano2002,Cirigliano2009} and this work.}
    \label{fig:OI_FM_nat}
\end{figure*}

Figure \ref{fig:OI_FM_nat} shows the overlap integrals weighted by natural abundance, in comparison with work from Kitano \textit{et al.} \cite{Kitano2002}.
The latter does not use natural abundance weighting but only displays one isotope per element.
Again, the main difference between the two datasets is in the deformed nuclei region.

From the overlap integrals, we can compute the conversion rate using Eq. \ref{eq:rate} for all the elements, and we obtain the sensitivity of different target materials normalized to the conversion rate in $^{27}$Al.
The relative sensitivity is defined as the conversion rate normalized by the muon capture rate \cite{Cirigliano2009}:

\begin{equation}
    \frac{R_{\mu e}(Z)}{R_{\mu e}(\textrm{Al})} = \frac{\frac{\Gamma_{conv}(Z)}{\Gamma_{capture}(Z)}}{\frac{\Gamma_{conv}(\textrm{Al})}{\Gamma_{capture}(\textrm{Al})}}
    \label{eq:sensitivity}
\end{equation}

\begin{figure*}[!ht]
    \centering
    \includegraphics[width=1.0\textwidth]{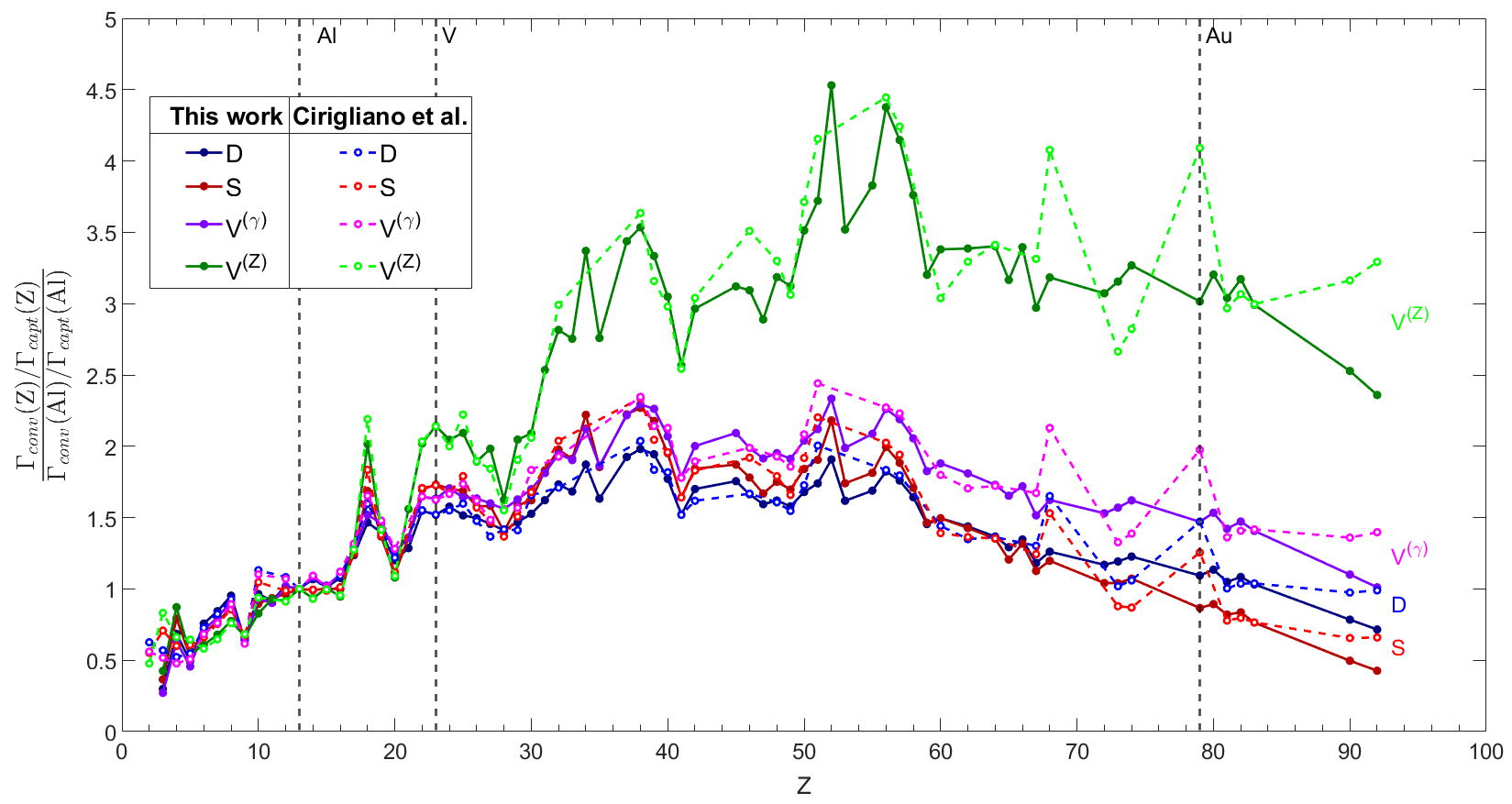}
    \caption{Comparison with of our results on sensitivity as a function of $Z$ with those of Cirigliano {\it et al.}\cite{Cirigliano2009}. The conversion rates are normalized to the muon capture rate and are relative to the rate for aluminum (see Eq. \ref{eq:sensitivity}).}
    \label{fig:sensitivity_comp_Cirigliano}
\end{figure*}

In Fig. \ref{fig:sensitivity_comp_Cirigliano}, our result is compared to that of Cirigliano \textit{et al.} \cite{Cirigliano2009}.

Four physics models are considered: dipole ($D$), scalar ($S$) and two vector type ($V^{(\gamma)}$ and $V^{(Z)}$). $V^{(\gamma)}$ describes the scenarios where the transition charge radius operator gives the dominant contribution to the CLFV Lagrangian, and  $V^{(Z)}$ describes the case where the dominant operator is assumed to be an effective $Z$-penguin~\cite{Ardu2023}.

\begin{figure*}[!ht]
    \centering
    \includegraphics[width=1.0\textwidth]{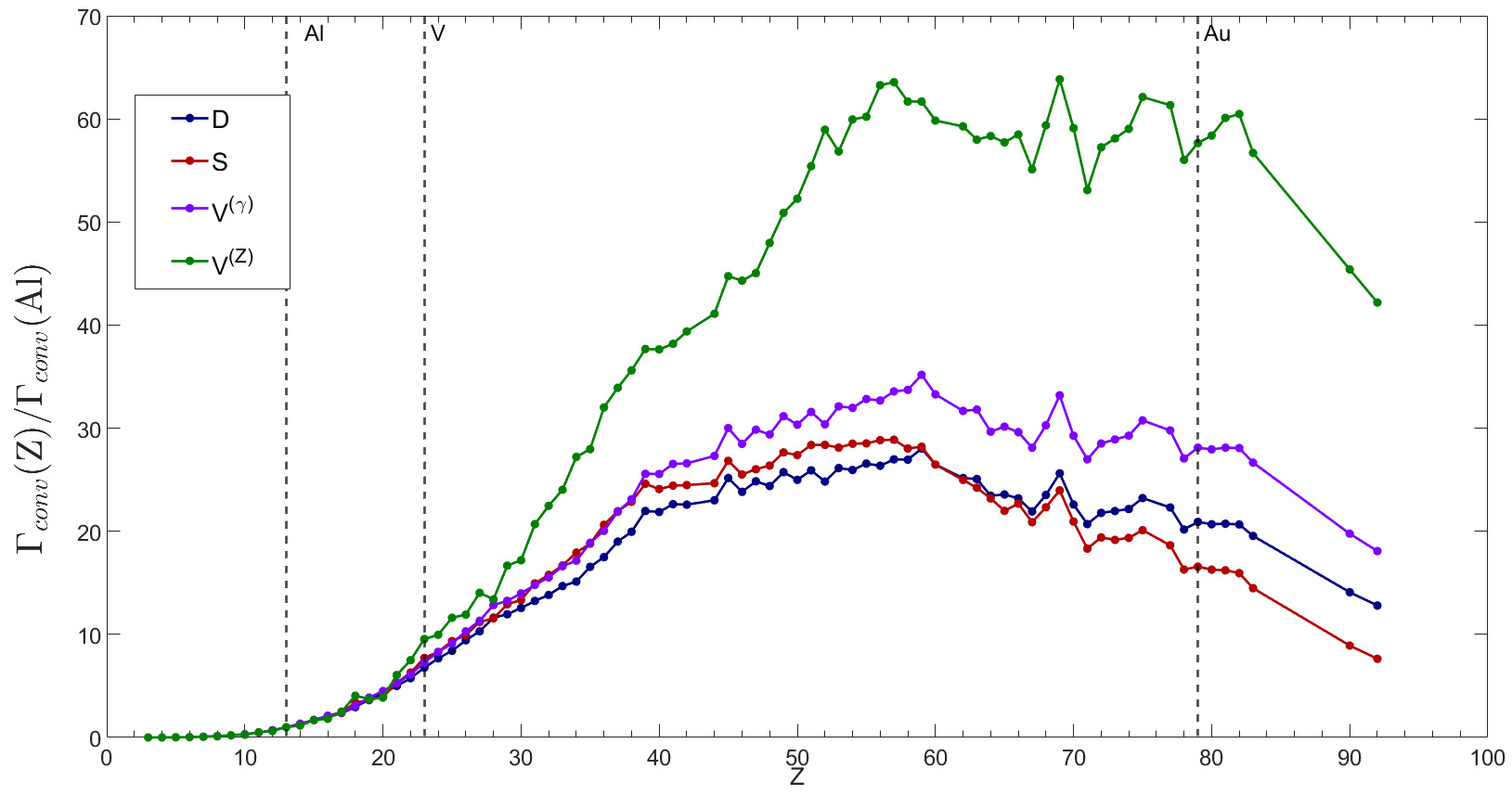}
    \caption{Our calculation of the experimental sensitivity of CLFV experiments as a function of atomic number. The conversion rates are not normalized to the muon capture rate, but are relative to the rate for aluminum (see Eq. \ref{eq:conv_rate}).}
    \label{fig:sensitivity_MZ_nat}
\end{figure*}

Figure \ref{fig:sensitivity_MZ_nat} removes the normalization relative to muon capture and shows the conversion rates in each material relative to that in $^{27}$Al. The objective is to separate the coherent muon-to-electron conversion process from the incoherent muon capture process. The motivation for this presentation of our results is detailed in Sec. \ref{sec:Normalization}.

\begin{table*}[!ht]
\centering
\caption{\label{table:Nd} Comparison of muonic X-ray and electron scattering determinations of two-parameter Fermi distributions in neodymium isotopes. $\beta$ comes directly from $B$(E2) measurement in \cite{Stelson1965}. The Barrett parameters values for Nd are $\alpha = 0.130$ and $k = 2.175$ \cite{Barrett1970}.}
\begin{tabular}{c| c c c c c c  } 
 \hline\hline
 \Tstrut Isotope & c (fm) & t (fm) & $\beta$ \cite{Stelson1965} &  {\it rms} radius (fm) & Barrett moment & Barrett radius $R_{k \alpha}$ (fm) \\ \hline
 \Tstrut & & & & Muonic X-ray\cite{Macagno1970} & & \\ \hline
 \Tstrut$^{142}$Nd  &  $5.80\pm0.03$ & $2.32\pm0.08$ & $0.104\pm0.02$ & $4.91\pm0.08$ & $15.99\pm 0.03$ & $6.27\pm0.01$ \\ 
 \Tstrut$^{144}$Nd & $5.85\pm0.03$ & $2.27\pm0.08$ & $0.123\pm0.02$ & $4.94\pm0.08$ & $16.11\pm 0.04$ & $6.31\pm0.01$ \\
 \Tstrut$^{146}$Nd & $5.82\pm0.03$ & $2.42\pm0.08$ & $0.151\pm0.02$ & $4.98\pm0.08$ & $16.26\pm 0.06$ & $6.34\pm0.02$ \\
 \Tstrut$^{148}$Nd & $5.84\pm0.03$ & $2.40\pm0.08$ & $0.197\pm0.04$ & $5.00\pm0.08$ & $16.38\pm 0.13$ & $6.38\pm0.03$ \\
 \Tstrut$^{150}$Nd & $5.86\pm0.03$ & $2.35\pm0.08$ & $0.279\pm0.005$ & $5.04\pm0.08$ & $16.58\pm 0.01$ & $6.43\pm0.01$ \\
 \hline
 \Tstrut & & & & Electron Scattering \cite{Heisenberg1971} & & \\ \hline
 \Tstrut $^{142}$Nd & $5.7045\pm0.03$ & $2.539\pm0.013$ & $0.104\pm0.02$ & $4.92\pm0.08$ & $15.98\pm 0.03$ & $6.27\pm0.01$ \\  
\Tstrut  $^{144}$Nd & $5.6634\pm0.03$ & $2.696\pm0.013$ & $0.123\pm0.02$ & $4.96\pm0.08$ & $16.11\pm 0.04$ & $6.31\pm0.01$ \\ 
 \Tstrut $^{146}$Nd & $5.6600\pm0.03$ & $2.760\pm0.013$ & $0.151\pm0.02$ & $4.99\pm0.08$ & $16.24\pm 0.06$ & $6.34\pm0.02$ \\
 \Tstrut $^{148}$Nd & $5.6871\pm0.03$ & $2.798\pm0.022$ & $0.197\pm0.04$ & $5.04\pm0.08$ & $16.47\pm 0.13$ & $6.40\pm0.03$ \\
 \Tstrut $^{150}$Nd & $5.7185\pm0.03$ & $2.861\pm0.031$ & $0.279\pm0.005$ & $5.13\pm0.08$ & $16.86\pm 0.01$ & $6.50\pm0.01$ \\
 \hline \hline             
\end{tabular}

\end{table*}

\section{\label{sec:Normalization}{Normalization and \texorpdfstring{$(Z,A)$}{(Z,A)} Dependence of the Conversion Rate}}

The conventional approach to the normalization of $\mu \to e$ conversion experiments, quoting the conversion rate (experimental limit or theory prediction) normalized to the measured rate of $\mu$ capture on a given nucleus, has been in place for more than seventy years. We will discuss the shortcomings of this convention and propose a revised presentation of $\mu \to e$ conversion results that addresses our concerns.

The first limit on  $\mu \to e$ conversion, by Lagarrigue and Peyrou~\cite{Lagarrigue1952}, using cosmic ray muons stopped in copper and tin targets in a Wilson cloud chamber, employed previous measurements of the muon lifetime (dominated by muon capture in this regime) in copper and antimony~\cite{Harrisson1951} as normalization. The first accelerator experiment, using a copper target~\cite{Steinberger1955}, also normalized to the (directly measured) rate of muon capture.

\par
The choice to normalize to the muon capture rate is not precisely analogous to the idea of a branching fraction (the number of decays into a particular mode, divided by all decays), which would be to divide the conversion rate in the field of a particular nucleus by all possible fates of the muon ($\mu\rightarrow e$ conversion, decay in orbit or nuclear capture). This normalization is essentially a historical convention initially codified by the work of Weinberg and Feinberg in 1959~\cite{Weinberg1959}. All results or predictions on muon-to-electron conversion have henceforth been presented in the form $R_{\mu \to e}$ (Eq.~\ref{eq:rmue}). 

\par
Compilations of the history of experimental limits on CLFV processes typically present the 90\% confidence level limits for decays and conversion on the same plot. This analogy ignores the fact that decay and conversion experiments are normalized differently. The decay limits are reported as true branching fractions, while the conversion rate limits are on the fraction of muon captures resulting in the production of a mono-energetic electron, which does not account for all fates of a muon in a muonic atom. Indeed, the lifetime of such a muon is determined in varying proportions by the conversion rate, a BSM process, by the nuclear capture rate, an incoherent Standard Model process, and by the lifetime of the decay-in-orbit muon, which is modified from the free decay rate by the atomic binding energy, the so-called Huff factor (0.993 for aluminum, 0.981 for titanium and 0.850 for gold) \cite{Suzuki1987}.

It is clear from these points that a revised approach is desirable. In particular, one would like to avoid the spurious introduction of nuclear effects into the presentation of a conversion result to the greatest extent possible to facilitate a conceptually cleaner comparison with results from decay experiments.

\par
Normalizing an experimental conversion result requires determination of the number of stopped muons within the sensitive time window of the experiment. This involves muon decay as well as nuclear muon capture.  In the upcoming Mu2e experiment downstream detectors count $2P-1S$ muonic X-rays or other transitions and delayed $\gamma$-rays from muon capture to infer the stopped muon rate. This measured rate is then used to normalize the conversion rate limit or observation.

\begin{table}[!ht]
\centering
\caption{\label{table:mucapAL} Muon capture rates on $^{27}$Al leading to Mg, Na and Ne final states (from~\cite{Measday2007}).}
\begin{tabular}{l c c c r} 
 \hline\hline
\Tstrut Reaction & Observed & Estimated & Missing & Total\\
   & $\gamma$-ray  & ground-state & yields & yield \\
\Tstrut     & yield  & transition &   &   \\
 \hline 
\Tstrut $^{27}$Al$(\mu^-\!,\nu)^{27}$Mg  & 10(1) & 0 & 3 & 13 \\
\Tstrut $^{27}$Al$(\mu^-\!,\nu n)^{26}$Mg  & 53(5) & 4 & 4 & 61 \\
\Tstrut $^{27}$Al$(\mu^-\!,\nu 2n)^{25}$Mg  &  7(1) & 3 & 2 & 12 \\
\Tstrut $^{27}$Al$(\mu^-\!,\nu 3n)^{24}$Mg  & 2 & 3 & 1 & 6 \\
\Tstrut $^{27}$Al$(\mu^-\!,\nu  p x n)^{26-23}$Na  & 2 & 2 & 1 & 5 \\
\Tstrut $^{27}$Al$(\mu^-\!,\nu  \alpha x n)^{23-21}$Ne  & 1 & 2 &0 & 3 \\
\Tstrut Total & 75(5) & 14 & 11 & 100 \\
 \hline \hline 
\end{tabular}

\end{table}

Table~\ref{table:mucapAL} lists the muon capture reactions in $^{27}$Al, which produce states of Mg, Na, and Ne \cite{Measday2007}. Note that only 13\% of captures result in the $^{27}$Mg ground state. The majority of captures result in neutron emission, with the nucleus left in either ground or excited states. 
The detailed calculation of the relevant nuclear matrix elements is quite complex, which is the reason the experimentally measured value of the total $\mu$ capture rate is used in deriving a conversion experiment's single event sensitivity. Knowledge of the number of stopped muons within the sensitive time window of the experiment is fundamental to normalization of the result.

\begin{figure}[!ht]
    \includegraphics[width=3.5in]{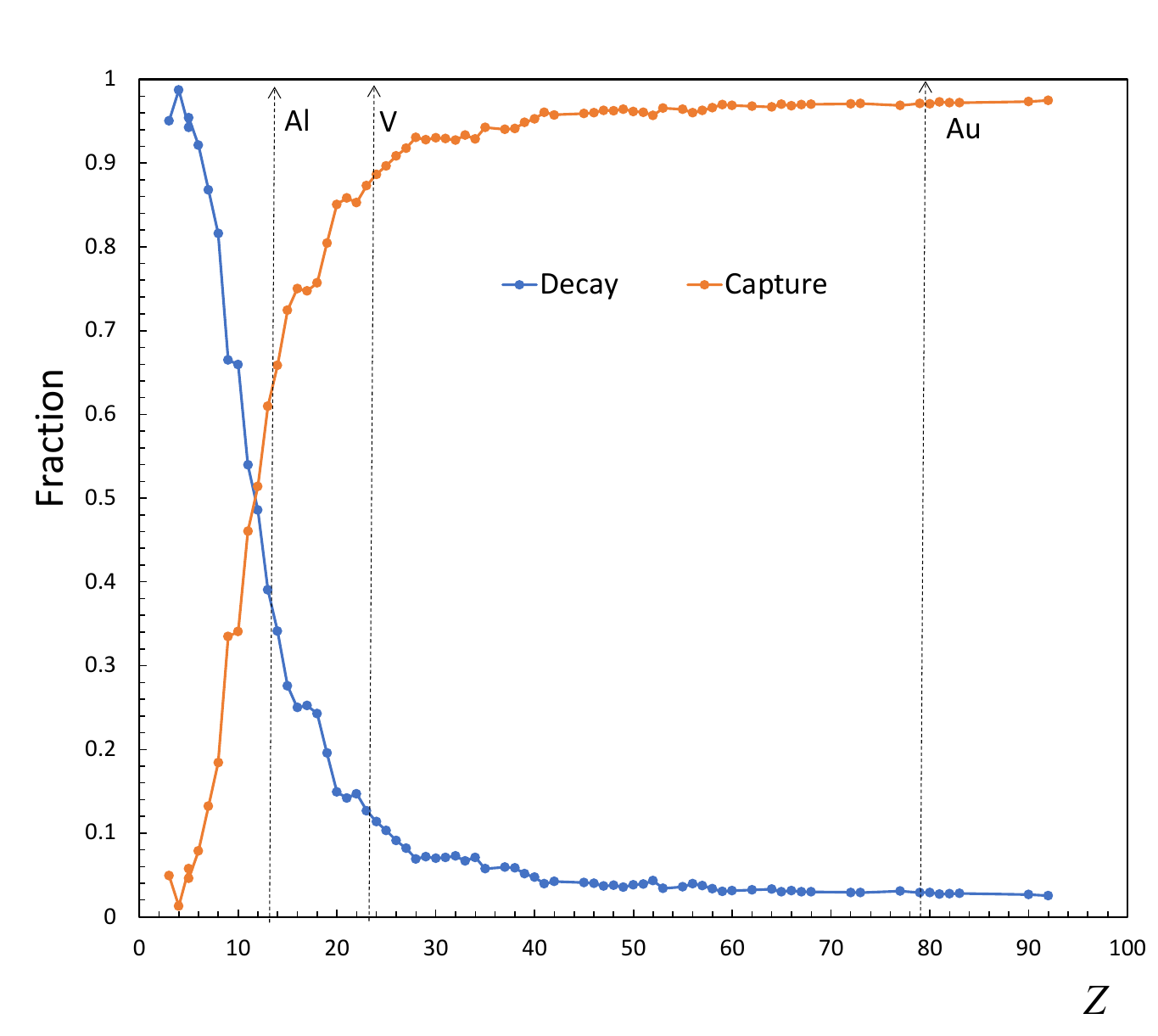}
    \caption{The fraction of muon decay in orbit in a muonic atom compared to the $\mu^-$ capture rate as a function of $Z$. Experimental values are taken from the Measday compilation~\cite{Measday2001}, a survey experiment~\cite{Suzuki1987} or a specific previous experiment (Ref. \cite{Measday2007} for $^{27}$Al).  }
    \label{fig:Suzuki}
\end{figure}

The experimental measurement of the total muon lifetime, with its associated uncertainties, thus unavoidably involves the calculation of the experimental efficiency and therefore the calculation of the $\mu \to e$ conversion rate. Since the overlap of the muon atomic wave function with the nuclear proton and neutron distribution influences the effective lifetime, the New Physics and the Standard Model nuclear physics are inextricably mixed: the measured rate (or limit on the rate) manifestly depends on the muon capture lifetime. The muon nuclear capture rate {\it grosso modo} follows Wheeler's \cite{Wheeler1949} $Z_{\rm eff}^4$ law, but in detail shows the effect of nuclear shell structure on nuclear size. The convention of quoting a ``capture fraction", thereby exaggerates the effect of nuclear shell model structure, complicating the isolation of New Physics from nuclear effects.

 \begin{figure*}[!ht]
    \centering
    \includegraphics[width=1.0\textwidth]
         {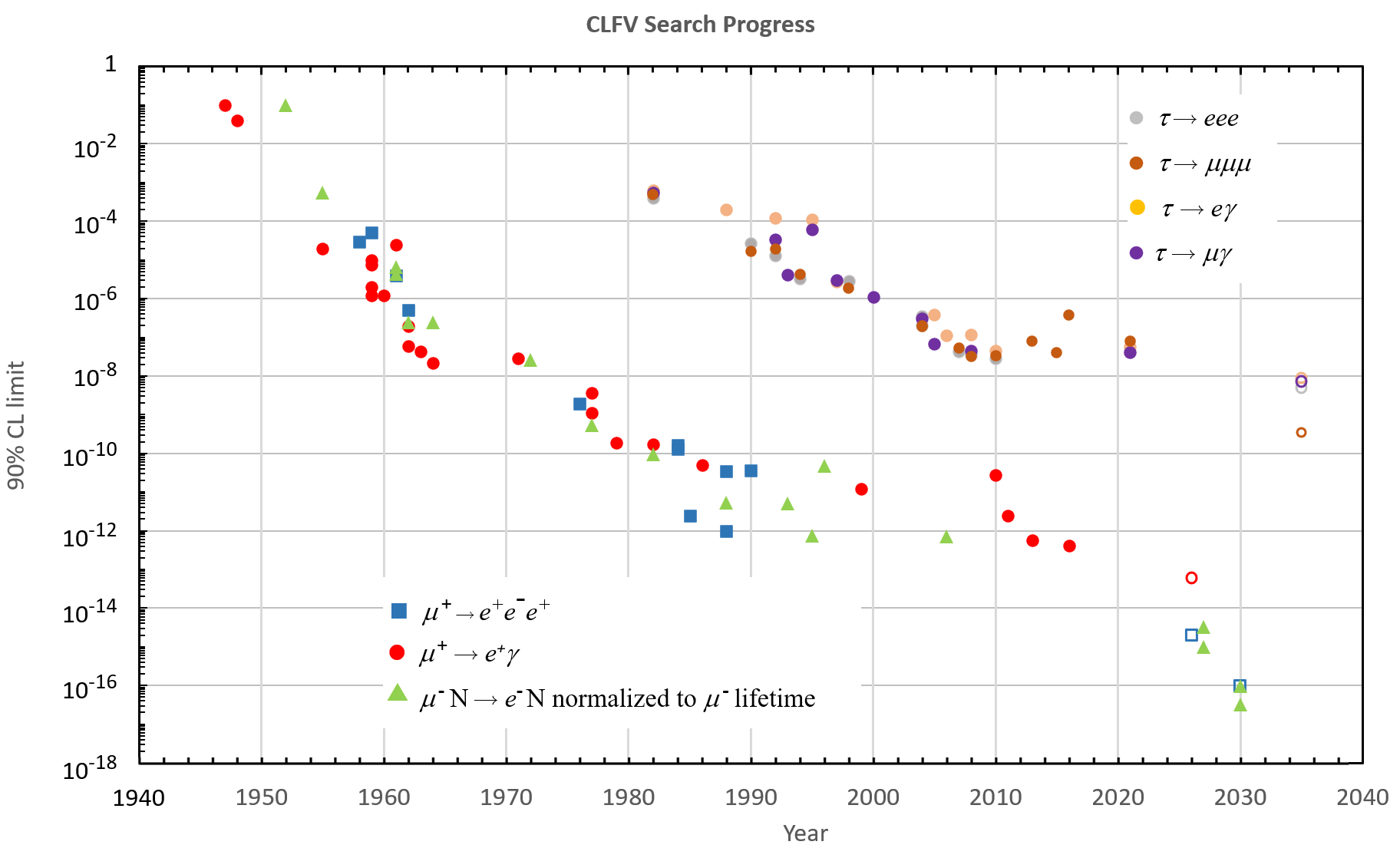}
        \caption{Chronology of 90\% confidence limits on $\mu\rightarrow e \gamma$, $\mu\rightarrow 3e$, $\mu$ to $e$ conversion, $\tau \to e \gamma$, $\tau \to \mu \gamma$ and $\tau \to 3e$, including predicted limits for current experiments. The  $\mu$ to $e$ conversion results are normalized to the total muon lifetime, not the heretofore conventional $\mu$ capture rate. Unfilled symbols represent the goals of experiments currently in preparation. Further improvement in search sensitivity is possible with new stopped muon beams under study for decades hence.}
        \label{fig:CLFV_History}
\end{figure*}

\bigskip

An explicit example can be found in equation (3) in the report of the SINDRUM II limit on the conversion rate in lead (Pb) \cite{Honecker1996}, where the denominator in the calculation of $B_{\mu e}$ is $f_{\rm capt} N_{\rm stop} \epsilon_{\rm tot}$. Since $f_{\rm capt}$ for Pb is 0.95, the normalization to $\mu$ capture is only slightly different from the normalization to muon stops. For lighter nuclei, however, for example, aluminum, $f_{\rm capt}$ is 0.61, so the change in normalization amounts to 64\%. Further, since the $\mu$ capture rate as a function of atomic number reflects nuclear shell structure as well the details of the low-lying levels of individual nuclei, and is primarily an incoherent process, its use introduces extraneous nuclear physics into consideration of the $Z$ dependence of the coherent New Physics process of $\mu$ to $e$ conversion.

\par
From a theoretical perspective, a model calculation of the rate of conversion effectively yields an absolute rate (more specifically, a rate characterized by $G_F^2$ and a mass-scale coupling factor). 
The conventional normalization involves a hybrid ratio of the calculated rate of the coherent conversion process divided by the experimental measurement of a partially incoherent muon capture process. This has not mattered in any practical sense up to this point, but makes conceptual comparisons with decay experiments unnecessarily difficult.  We should remove this inconsistency before we begin to compare measured $\mu \to e$ conversion rates for different nuclei to explore the Lorentz structure of the conversion process.

\begin{table}[!ht]
\centering
\caption{\label{table:renorm} Atomic number-dependent adjustment of $\mu \to e$ conversion limits to normalization to muon stops rather than captures.\smallskip}
\begin{tabular}{c|c|c|c|c} 
 \hline\hline
\Tstrut  &  &  & 90\% CL Limit & 90\% CL Limit\\
 Year  & Reference  & Nucleus & Normalized to & Normalized to \\
\Tstrut     &   &  & $\mu$ Capture & $\mu$ Stops \\
 \hline 
1952 \Tstrutt & \cite{Lagarrigue1952} & Sn, Sb & $1.0 \times 10^{-1}$ & $1.0 \times 10^{-1}$\\
1955 \Tstrut& \cite{Steinberger1955}& Cu & $5.0 \times 10^{-4}$ & $5.4 \times 10^{-4}$\\
1961 \Tstrut& \cite{Sard1961}& Cu & $4.0 \times 10^{-6}$ & $4.3 \times 10^{-6}$\\
1961 \Tstrut& \cite{Conversi1961} & Cu &$ 5.9 \times 10^{-6}$ & $6.4 \times 10^{-6}$\\
1962 \Tstrut& \cite{Conforto1962} & Cu &$ 2.2 \times 10^{-7}$ & $2.4 \times 10^{-7}$\\
1964 \Tstrut& \cite{Bartley1964} & Cu &$ 2.2 \times 10^{-7}$ & $2.4 \times 10^{-7}$\\
1972 \Tstrut & \cite{Bryman1972} & Cu &$ 2.6 \times 10^{-8}$ & $2.8 \times 10^{-8}$\\
1977 \Tstrut& \cite{Badertscher1977} & S & $4.0 \times 10^{-10}$ & $5.3 \times 10^{-10}$\\
1982 \Tstrut& \cite{Badertscher1982} & S & $ 7.0 \times 10^{-11}$ & $9.3 \times 10^{-11}$\\
1988 \Tstrut& \cite{Ahmad1988} & Ti & $ 4.6 \times 10^{-12}$ & $5.4 \times 10^{-12}$\\
1993 \Tstrut& \cite{Dohmen1993} & Ti & $ 4.3 \times 10^{-12}$ & $5.0 \times 10^{-12}$\\
1995 \Tstrut& \cite{Eggli1995} & Ti &$ 6.5 \times 10^{-13}$ & $7.6 \times 10^{-13}$\\
1996 \Tstrut& \cite{Honecker1996} & Pb & $ 4.6 \times 10^{-11}$ & $4.7 \times 10^{-11}$\\
2006 \Tstrut& \cite{SINDRUMII2006} & Au &$ 7.0 \times 10^{-13}$ & $7.2 \times 10^{-13}$\\
 \hline \hline
\end{tabular}

\end{table}

Crivellin {\it et al.} \cite{Crivellin2017} write the conversion rate, \textbf{not} the conversion ratio, as
\begin{equation}
    \Gamma_{conv} = \,\frac{{m_\mu ^5}}{{4{{\rm{\Lambda }}^4}}}| eC_L^D{D_N} + 4({G_F}{m_\mu }{m_p}\tilde C_{(p)}^{SL}S_N^{(p)}
    \label{eq:conv_rate}
\end{equation}

\vspace{-20pt}
\begin{equation*}
\begin{gathered}
+ \  \tilde C_{(p)}^{VR}V_N^{(p)} + \,p \to n) | ^2 + L \leftrightarrow R, 
\end{gathered}
\end{equation*}
\par\noindent
involving the proton and neutron fields, but otherwise similar to the muon decay rates ($C$ are dimensionless Wilson coefficients and $\Lambda$ is the effective mass scale). 

The historical normalization of this process to the muon capture rate conflates the actual sensitivity of a conversion experiment with the nuclear physics of muon capture. In the background-free case the experimental sensitivity depends only on the number of conversion electrons in the signal window over the sensitive time of the experiment, which is also true of a rare decay experiment. Why then should we insert an extraneous muon capture factor into the many theoretical comparisons of conversion {\it vs.} decay sensitivity against particular models? 

 The limits normalized to muon stops and muon capture are listed in Table~\ref{table:renorm}. Fig.~\ref{fig:sensitivity_MZ_nat} presents our calculation of the $(Z,A)$ dependence of the CLFV matrix elements without the muon capture normalization, demonstrating the much-reduced extraneous structure heretofore introduced by normalization to the muon capture rate.
 
 Figure~\ref{fig:CLFV_History} shows the existing $\mu\rightarrow e$ conversion and decay experimental limits. Here we normalize the conversion results to muon stops, removing the historical muon capture normalization, making the comparison with decay experiments more consistent.

\par\noindent

\section{\label{sec:target_optimization}{Possible Target Considerations}}

Coherent muon-to-electron conversion in a nucleus results in the emission of a mono-energetic electron with an energy $E_{\mu e}$ that is nucleus-dependent (104.97 MeV for aluminum (Al)), with radiative corrections calculated in Ref.~\cite{Szafron2017}. The dominant intrinsic background in a conversion search is electrons from decay-in-orbit. The momentum spectrum of these electrons has a long recoil tail with significant radiative corrections \cite{Czarnecki2011,Szafron2015} and an endpoint close to the conversion electron signal. Another background source are electrons from radiative pion capture, produced when pions in the muon beamline stop in the target. These pions are captured into the orbit of the nucleus, resulting in the emission of a photon, $\pi^{-} + N(A,Z) \rightarrow \gamma^{(*)} + N(A, Z-1)$, followed by an asymmetric ($\gamma \rightarrow e^+e^-$) conversion producing electrons with energies nearly up to the charged pion mass ($139$ MeV/c$^{2}$). The upcoming Mu2e and COMET experiments mitigate pion backgrounds by using pulsed beams and by exploiting the short pion lifetime ($\sim$26 ns at rest), delaying the data-taking window by a few hundred ns. Since the time distribution of the conversion electron depends on the mean lifetime of the muonic atom and is nucleus-dependent (see Table ~\ref{table:elements}), this has implications for which target materials can be studied. Cosmic ray-induced backgrounds must also be vetoed, but have a small dependence on stopping target design \cite{SU2020}.

Table ~\ref{table:elements} details four potential target materials: titanium, vanadium, lithium, and gold. Figure~\ref{fig:Suzuki} shows the variation of the average capture and decay fractions with atomic number. Heavier nuclei have smaller decay fractions and larger capture fractions, which is beneficial, as it results in fewer decay background electrons for the same number of stopped muons.

\begin{table}[!ht]
\centering
\caption{\label{table:elements}Elements  and their characteristics (atomic number ($Z$), density ($\rho$), lifetime ($\tau_{mean}$), and relative average fraction of decay ($f_{dec.}$) to capture ($f_{capt.}$) along with their naturally occurring abundance ($Ab.$) for given isotopes ($I.$), and their nuclear spins ($s$)). The experimental values for capture and decay fractions are taken from the Measday compilation~\cite{Measday2001}, a survey experiment~\cite{Suzuki1987} or a specific previous experiment (Ref. \cite{Measday2007} for $^{27}$Al).\smallskip}
\begin{tabular}{c| c cc  c  c | c  c c c  } 
 \hline\hline
\Tstrut Element & $Z$ & $\rho$  & $\tau_{mean}$   &$f_{dec.}$ & $f_{capt.}$  & $I.$ & $Ab.$ & $s$ \\
 &  &  [g/cm$^{3}]$ &  [ns] & [$\%$] & [$\%$]  & &[$\%$] &\\
 \hline
\Tstrut Li & 3 &  0.534  & 2175&99 &1 & \Tstrut $^{6}$Li&7.6&1 \\
&&&&&&\Tstrut $^{7}$Li&92.4& {\scriptsize 3/2}\\
 \hline 
\Tstrut Al & 13 & 2.11  &864& 39 & 61 &\Tstrut  $^{27}$Al&100& {\scriptsize 5/2}\\
 \hline
\Tstrut Ti & 22&4.51&329&15 &85 &\Tstrut  $^{46}$Ti&8.25& 0\\
&&&&&&\Tstrut $^{47}$Ti&7.44& {\scriptsize 5/2}\\
&&&&&&\Tstrut $^{48}$Ti&73.72& 0\\
&&&&&&\Tstrut $^{49}$Ti&5.41& {\scriptsize 7/2}\\\
&&&&&&\Tstrut $^{50}$Ti&5.18& 0\\
 \hline 
\Tstrut V & 23&6.11&284&13 &87 &\Tstrut  $^{50}$V&0.25& 6\\
&&&&&&\Tstrut $^{51}$V&99.75& {\scriptsize 7/2}\\
 \hline 
\Tstrut Au &79 &19.3&73&3 &97 &\Tstrut  $^{197}$Au&100& {\scriptsize 3/2}\\ \hline\hline

\end{tabular}

\end{table}

We have considered only coherent contributions from dipole, scalar, and vector interaction, as these conversion rates are enhanced by $A^2$. Spin-dependent (SD) contributions, which do not have this coherent enhancement, have been studied in Refs.~\cite{Davidson2018,Davidson2019,noel,Haxton2024}.  The dipole, scalar, and vector operators contribute to the Spin-independent (SI) rate, while axial, tensor, and pseudoscalar operators contribute to the SD rate.  The SD rate depends on the distribution of spin in the nucleus \cite{Cirigliano2017}, and therefore requires detailed modeling of the target nucleus. In most models, SI conversion dominates due to this $A^{2}$ enhancement, but this is not true in all models. For example, Ref.~\cite{Fuyuto2023} describes how the SD interaction induced by ALP$-\pi^{0}$ mixing is the leading contribution to $\mu \to e$ conversion in models where axion-like particles (ALPs) induce the conversion. Consequently, the spin of the target nuclei is important in unraveling the Lorentz structure of the New Physics.

From a single Al conversion measurement, we will not know the form of the coupling. Figs.~\ref{fig:sensitivity_comp_Cirigliano} and \ref{fig:sensitivity_MZ_nat} show that if conversion is measured in Al, both titanium (Ti) and vanadium (V) have the advantage of providing good separation in the relative conversion rates produced by the dipole, scalar, and vector interactions. This makes them useful for determining the type of physics responsible. From a practical point of view, Table~\ref{table:elements} shows that the mean lifetime of a muonic Ti or V atom is not so short that pion-induced backgrounds would overwhelm the conversion signal. Ti and V have smaller average decay fractions than Al, meaning lower decay backgrounds for the same number of stopped muons. For these reasons Ti and V are viable target choices for the next generation of conversion measurements. Natural titanium has several stable isotopes with different spins and separated isotopes of Ti are not readily available in the quantities needed for a target of $\sim 200$ g. Employing a natural Ti target introduces odd-nucleon isotopes which could produce a non-zero SD response. These SD contributions could be distinguished from the coherent SI response \cite{Haxton2022}, however, this mixing of isotopes complicates the unraveling of SD and SI conversion rates. Vanadium is an attractive alternative; it provides the same physics benefits (increased survival fraction, similar muonic lifetime, {\it etc.}) with the advantage that it has a single stable isotope, $^{51}$V (spin $\frac{7}{2}$), with $>$~99$\%$ abundance.

 Figures~\ref{fig:sensitivity_comp_Cirigliano} and \ref{fig:sensitivity_MZ_nat} present gold (Au) as an obvious choice. Au has several additional physics benefits, Table~\ref{table:elements} shows that there is one stable isotope and that the relative fraction of decay to capture in Au is small, meaning on average 3 $\%$ of muons will undergo decay, a significant reduction from 39 $\%$ in Al. Unfortunately, Au has a mean muonic atom lifetime of just 73 ns so it would not be feasible to use a Au target in a Mu2e or COMET style design due to the enforced time selection required to remove pion-induced backgrounds. If pion contamination could be eliminated using a Fixed Field Alternating-gradient machine such as in the proposed Advanced Muon Facility \cite{Aoki2022}, measuring conversion in Au could be feasible.

Comparing conversion in light targets with very different neutron-to-proton ratios (for example lithium (Li) and Al)  could allow us to distinguish operators involving neutrons from those involving protons \cite{Davidson2019}. Additionally, Ref. \cite{Davidson2018} suggests that a conversion measurement in lighter nuclei, such as Li, would be of interest for detecting SD conversion since the SD rate is relatively suppressed by $1/A^{2}$ compared to the SI rate. As a result, the ratio $\omega_{conv, SD}/\omega_{conv, SI}$ is larger in lighter nuclei.  A Li foil-style target is feasible \cite{BNL-CERN-Syracuse-Yale1977}. However, the disadvantage here is that the muonic lifetime in Li is close to the free muon lifetime. As a result, the survival fraction is small and the relative decay fraction would be much larger than in Al, meaning more decay background for the same number of stopped muons. Since Li has a much lower density, $\sim$ 4 times less than Al, stopping the same number of muons would require a much larger target volume and due to the much smaller survival fraction, reaching the same sensitivity as aluminum would require a much longer running time.

All the target materials presented here have physics advantages and the specific target chosen will depend on, and influence, the experimental design of any future facility, which is beyond the scope of this paper.

\section{\label{sec:Conclusions}{Conclusions}}

To conclude, we have presented a new calculation of the $(Z,A)$ dependence of coherent muon-to-electron conversion in 236 isotopes, significantly extending the previous studies of Kitano {\it et al.}~\cite{Kitano2002}  and Cirigliano {\it et al.}~\cite{Cirigliano2009}. Our approach has several key improvements over these previous treatments:
\begin{itemize}
    \item We have included the effect of permanent quadrupole deformation on the CLFV matrix elements, using the method of Barrett moments to add the substantial catalog of muonic X-ray nuclear size and shape determinations of the charge distribution to the electron scattering results used in previous calculations.
    \item Rather than using neutron distributions equal to charge distributions, as in the previous work, we have employed the DRHBc theory for even-even nuclei to calculate neutron-related matrix elements. This takes into account the quadrupole deformation of the neutron distributions as well as the fact that neutrons are in general in different shell model orbits than protons.
    \item  We present the resulting CLFV overlap integrals for stable isotopes with greater than 1\% abundance as well as results weighted for natural abundance.
    \item The resulting conversion rates differ from previous calculations, particularly in the region of large permanent quadrupole deformation.
\end{itemize}

Finally, we propose a revised normalization for $\mu \to e$ conversion results, quoting the measured conversion rate (or limit thereon) directly, instead of presenting the conversion rate divided by the muon capture rate, which is generally not measured in the same experiment. 

Our sensitivity plots are presented in this form. Note the reduction in the scatter of points, which is largely due to the removal of the additional shell model structure observed in the incoherent muon capture process.

\section*{Acknowledgements}
This work was supported in part by DOE Grant DE‐SC0011925. It was initiated at the Aspen Center for Physics, which is supported by National Science Foundation grant PHY-2210452.

We acknowledge helpful conversations with Wick Haxton, Julian Heeck, Natalia Oreshkina, Ryan Plestid, Robert Szafron and Petr Vogel, as well as the careful reading of the text by Bertrand Echenard, James Miller, Ryan Plestid and Frank Porter.

\renewcommand{\thefootnote}{\alph{footnote}}

\clearpage
\newpage

\appendix

\onecolumngrid
\section{\label{sec:long_table} Table of Overlap Integrals}

\vspace{-8pt}
The results obtained using the method described herein for the 236 stable isotopes with natural abundance $> 1\%$.

The description of the columns is as follows:
\begin{center}
\begin{tabular}{ll}
    \textbf{Isotope:} Isotope and nucleon number \phantom{aaaa} & {$\boldsymbol\beta$:}
    Permanent quadrupole deformation\\
    {$\boldsymbol{Z}:$ Proton number} & $\boldsymbol{\alpha}$ and $\boldsymbol{k}:$ Barrett parameters (Sec. \ref{sec:dataset}) \\
    $\boldsymbol{N}:$ Neutron number &  $\boldsymbol{R_{k \alpha}}:$ Barrett radius [fm]  (Sec. \ref{sec:dataset})\\
    $\boldsymbol{c}:$ Radius at half maximum [fm] & $\boldsymbol{D, S(p), V(p), S(n), V(n)}:$ \\ 
    $\boldsymbol{t_{eq}}:$ Equivalent skin thickness [fm] (Sec. \ref{sec:deformation}) & \phantom{aaaa} Overlap integrals [$m_{\mu}^{5/2}$] (Sec. \ref{sec:theory}) \\
\end{tabular}
\end{center}

\onecolumngrid
\renewcommand{\thefootnote}{\alph{footnote}}
\setcounter{table}{0}
\renewcommand{\thetable}{A.\arabic{table}}

\begin{longtable}{|c|c|c|c|c|c|c|c|c|c|}
\caption{\label{table:undeformed}Table of overlap integrals for nuclei assumed spherical}\\
\noalign{\vspace{1.5em}}
    \hline
    \Tstrut Isotope & $Z$ & $N$ & $c$ & $t$ & $D$ & $S(p)$ & $V(p)$ & $S(n)$ & $V(n)$ \\ \hline
    \Tstrut $^{6}$Li\footnotemark[1] & 3 & 3 & 2.13 & 2.3 & 0.00131 & 0.00055 & 0.00055 & 0.00055 & 0.00055 \\ \hline
    \Tstrut $^{7}$Li\footnotemark[1] & 3 & 4 & 1.75 & 2.3 & 0.00137 & 0.00058 & 0.00058 & 0.00077 & 0.00078 \\ \hline
    \Tstrut $^{9}$Be & 4 & 5 & 1.789 & 2.3 & 0.00277 & 0.00117 & 0.00118 & 0.00146 & 0.00147 \\ \hline
    \Tstrut $^{10}$B & 5 & 5 & 1.928 & 2.3 & 0.00468 & 0.00198 & 0.002 & 0.00198 & 0.002 \\ \hline
    \Tstrut $^{11}$B\footnotemark[1] & 5 & 6 & 1.82 & 2.3 & 0.00475 & 0.00201 & 0.00203 & 0.00241 & 0.00244 \\ \hline
    \Tstrut $^{12}$C & 6 & 6 & 2.001 & 2.3 & 0.0072 & 0.00306 & 0.0031 & 0.00306 & 0.0031 \\ \hline
    \Tstrut $^{13}$C & 6 & 7 & 1.996 & 2.3 & 0.0072 & 0.00306 & 0.0031 & 0.00357 & 0.00362 \\ \hline
    \Tstrut $^{14}$N & 7 & 7 & 2.151 & 2.3 & 0.01018 & 0.00434 & 0.00441 & 0.00434 & 0.00441 \\ \hline
    \Tstrut $^{16}$O & 8 & 8 & 2.413 & 2.3 & 0.01338 & 0.00571 & 0.00582 & 0.00571 & 0.00582 \\ \hline
    \Tstrut $^{19}$F & 9 & 10 & 2.776 & 2.3 & 0.01644 & 0.007 & 0.00717 & 0.00778 & 0.00797 \\ \hline
    \Tstrut $^{20}$Ne & 10 & 10 & 2.959 & 2.3 & 0.02016 & 0.00859 & 0.00883 & 0.00859 & 0.00883 \\ \hline
    \Tstrut $^{22}$Ne & 10 & 12 & 2.871 & 2.3 & 0.02057 & 0.00877 & 0.00901 & 0.01053 & 0.01081 \\ \hline
    \Tstrut $^{23}$Na & 11 & 12 & 2.939 & 2.3 & 0.02521 & 0.01077 & 0.0111 & 0.01175 & 0.01211 \\ \hline
    \Tstrut $^{24}$Mg & 12 & 12 & 3.045 & 2.3 & 0.02996 & 0.01281 & 0.01325 & 0.01281 & 0.01325 \\ \hline
    \Tstrut $^{25}$Mg & 12 & 13 & 2.998 & 2.3 & 0.03031 & 0.01297 & 0.0134 & 0.01405 & 0.01452 \\ \hline
    \Tstrut $^{26}$Mg & 12 & 14 & 3.007 & 2.3 & 0.03024 & 0.01294 & 0.01337 & 0.01509 & 0.0156 \\ \hline
    \Tstrut $^{27}$Al & 13 & 14 & 3.055 & 2.3 & 0.03579 & 0.01533 & 0.0159 & 0.01651 & 0.01712 \\ \hline
    \Tstrut $^{28}$Si & 14 & 14 & 3.154 & 2.3 & 0.04118 & 0.01764 & 0.01837 & 0.01764 & 0.01837 \\ \hline
    \Tstrut $^{29}$Si & 14 & 15 & 3.148 & 2.3 & 0.04124 & 0.01767 & 0.0184 & 0.01893 & 0.01971 \\ \hline
    \Tstrut $^{30}$Si & 14 & 16 & 3.172 & 2.3 & 0.04099 & 0.01755 & 0.01828 & 0.02006 & 0.02089 \\ \hline
    \Tstrut $^{31}$P & 15 & 16 & 3.265 & 2.3 & 0.04655 & 0.01994 & 0.02084 & 0.02127 & 0.02223 \\ \hline
    \Tstrut $^{32}$S & 16 & 16 & 3.382 & 2.3 & 0.05187 & 0.0222 & 0.02331 & 0.0222 & 0.02331 \\ \hline
    \Tstrut $^{34}$S & 16 & 18 & 3.418 & 2.3 & 0.05134 & 0.02196 & 0.02306 & 0.0247 & 0.02594 \\ \hline
    \Tstrut $^{35}$Cl\footnotemark[2] & 17 & 18 & 3.344 & 2.6718 & 0.05534 & 0.02364 & 0.02493 & 0.02503 & 0.0264 \\ \hline
    \Tstrut $^{37}$Cl\footnotemark[2] & 17 & 20 & 3.38 & 2.6762 & 0.05473 & 0.02336 & 0.02464 & 0.02748 & 0.02899 \\ \hline
    \Tstrut $^{40}$Ar & 18 & 22 & 3.642 & 2.3 & 0.0617 & 0.02632 & 0.0279 & 0.03217 & 0.03409 \\ \hline
    \Tstrut $^{39}$K & 19 & 20 & 3.654 & 2.3 & 0.06886 & 0.0294 & 0.03126 & 0.03095 & 0.03291 \\ \hline
    \Tstrut $^{41}$K & 19 & 22 & 3.682 & 2.3 & 0.06825 & 0.02912 & 0.03098 & 0.03372 & 0.03587 \\ \hline
    \Tstrut $^{40}$Ca & 20 & 20 & 3.722 & 2.3 & 0.07495 & 0.03198 & 0.03414 & 0.03198 & 0.03414 \\ \hline
    \Tstrut $^{44}$Ca & 20 & 24 & 3.784 & 2.3 & 0.0734 & 0.03126 & 0.03341 & 0.03752 & 0.0401 \\ \hline
    \Tstrut $^{45}$Sc & 21 & 24 & 3.828 & 2.3 & 0.07996 & 0.03404 & 0.03653 & 0.03891 & 0.04175 \\ \hline
    \Tstrut $^{46}$Ti & 22 & 24 & 3.92 & 2.3 & 0.0851 & 0.03616 & 0.03899 & 0.03945 & 0.04254 \\ \hline
    \Tstrut $^{47}$Ti & 22 & 25 & 3.904 & 2.3 & 0.08559 & 0.03639 & 0.03923 & 0.04135 & 0.04457 \\ \hline
    \Tstrut $^{48}$Ti & 22 & 26 & 3.898 & 2.3 & 0.08576 & 0.03647 & 0.0393 & 0.0431 & 0.04645 \\ \hline
    \Tstrut $^{49}$Ti & 22 & 27 & 3.871 & 2.3 & 0.08661 & 0.03686 & 0.03971 & 0.04524 & 0.04873 \\ \hline
    \Tstrut $^{50}$Ti & 22 & 28 & 3.866 & 2.3 & 0.08675 & 0.03693 & 0.03978 & 0.047 & 0.05062 \\ \hline
    \Tstrut $^{51}$V & 23 & 28 & 3.91 & 2.3 & 0.09342 & 0.03973 & 0.04298 & 0.04837 & 0.05232 \\ \hline
    \Tstrut $^{50}$Cr & 24 & 26 & 4.004 & 2.3 & 0.09829 & 0.04171 & 0.04534 & 0.04518 & 0.04912 \\ \hline
    \Tstrut Isotope & $Z$ & $N$ & $c$ & $t$ & $D$ & $S(p)$ & $V(p)$ & $S(n)$ & $V(n)$ \\ \hline
    \Tstrut $^{52}$Cr & 24 & 28 & 3.974 & 2.3 & 0.09936 & 0.0422 & 0.04585 & 0.04924 & 0.05349 \\ \hline
    \Tstrut $^{53}$Cr & 24 & 29 & 4 & 2.3 & 0.09841 & 0.04176 & 0.0454 & 0.05046 & 0.05486 \\ \hline
    \Tstrut $^{54}$Cr & 24 & 30 & 4.042 & 2.3 & 0.09687 & 0.04105 & 0.04466 & 0.05131 & 0.05583 \\ \hline
    \Tstrut $^{55}$Mn & 25 & 30 & 4.073 & 2.3 & 0.1038 & 0.04396 & 0.04802 & 0.05276 & 0.05763 \\ \hline
    \Tstrut $^{54}$Fe & 26 & 28 & 4.055 & 2.3 & 0.11286 & 0.04785 & 0.05242 & 0.05153 & 0.05645 \\ \hline
    \Tstrut $^{56}$Fe & 26 & 30 & 4.12 & 2.3 & 0.11005 & 0.04655 & 0.05107 & 0.05371 & 0.05893 \\ \hline
    \Tstrut $^{57}$Fe & 26 & 31 & 4.144 & 2.3 & 0.109 & 0.04607 & 0.05057 & 0.05493 & 0.0603 \\ \hline
    \Tstrut $^{59}$Co & 27 & 32 & 4.196 & 2.3 & 0.11488 & 0.04847 & 0.05346 & 0.05745 & 0.06335 \\ \hline
    \Tstrut $^{58}$Ni & 28 & 30 & 4.177 & 2.3 & 0.12412 & 0.05242 & 0.05797 & 0.05616 & 0.06211 \\ \hline
    \Tstrut $^{60}$Ni & 28 & 32 & 4.233 & 2.3 & 0.12136 & 0.05115 & 0.05664 & 0.05845 & 0.06474 \\ \hline
    \Tstrut $^{61}$Ni & 28 & 33 & 4.249 & 2.3 & 0.12056 & 0.05078 & 0.05626 & 0.05984 & 0.06631 \\ \hline
    \Tstrut $^{62}$Ni & 28 & 34 & 4.277 & 2.3 & 0.11922 & 0.05016 & 0.05561 & 0.0609 & 0.06753 \\ \hline
    \Tstrut $^{63}$Cu & 29 & 34 & 4.338 & 2.3 & 0.12419 & 0.05212 & 0.05809 & 0.06111 & 0.0681 \\ \hline
    \Tstrut $^{65}$Cu & 29 & 36 & 4.367 & 2.3 & 0.12268 & 0.05143 & 0.05736 & 0.06384 & 0.0712 \\ \hline
    \Tstrut $^{64}$Zn & 30 & 34 & 4.405 & 2.3 & 0.12865 & 0.05385 & 0.06033 & 0.06103 & 0.06837 \\ \hline
    \Tstrut $^{66}$Zn & 30 & 36 & 4.435 & 2.3 & 0.12699 & 0.05308 & 0.05952 & 0.0637 & 0.07143 \\ \hline
    \Tstrut $^{67}$Zn\footnotemark[3] & 30 & 37 & 4.408 & 2.3 & 0.12846 & 0.05376 & 0.06023 & 0.0663 & 0.07429 \\ \hline
    \Tstrut $^{68}$Zn & 30 & 38 & 4.46 & 2.3 & 0.12563 & 0.05246 & 0.05887 & 0.06645 & 0.07457 \\ \hline
    \Tstrut $^{69}$Ga & 31 & 38 & 4.507 & 2.3 & 0.13075 & 0.05448 & 0.06143 & 0.06678 & 0.0753 \\ \hline
    \Tstrut $^{71}$Ga & 31 & 40 & 4.528 & 2.3 & 0.12952 & 0.05392 & 0.06084 & 0.06957 & 0.0785 \\ \hline
    \Tstrut $^{70}$Ge & 32 & 38 & 4.569 & 2.3 & 0.13483 & 0.05601 & 0.0635 & 0.06651 & 0.07541 \\ \hline
    \Tstrut $^{72}$Ge & 32 & 40 & 4.593 & 2.3 & 0.13338 & 0.05534 & 0.0628 & 0.06918 & 0.0785 \\ \hline
    \Tstrut $^{73}$Ge & 32 & 41 & 4.605 & 2.3 & 0.13263 & 0.055 & 0.06244 & 0.07047 & 0.08 \\ \hline
    \Tstrut $^{74}$Ge & 32 & 42 & 4.619 & 2.3 & 0.13182 & 0.05463 & 0.06204 & 0.07171 & 0.08143 \\ \hline
    \Tstrut $^{76}$Ge & 32 & 44 & 4.629 & 2.3 & 0.13117 & 0.05433 & 0.06173 & 0.07471 & 0.08488 \\ \hline
    \Tstrut $^{75}$As & 33 & 42 & 4.653 & 2.3 & 0.13719 & 0.05675 & 0.06475 & 0.07222 & 0.08241 \\ \hline
    \Tstrut $^{76}$Se & 34 & 42 & 4.716 & 2.3 & 0.14051 & 0.05791 & 0.06647 & 0.07154 & 0.08211 \\ \hline
    \Tstrut $^{77}$Se & 34 & 43 & 4.716 & 2.3 & 0.1405 & 0.05791 & 0.06647 & 0.07324 & 0.08406 \\ \hline
    \Tstrut $^{78}$Se & 34 & 44 & 4.718 & 2.3 & 0.14036 & 0.05785 & 0.0664 & 0.07486 & 0.08593 \\ \hline
    \Tstrut $^{80}$Se & 34 & 46 & 4.718 & 2.3 & 0.1404 & 0.05787 & 0.06642 & 0.07829 & 0.08986 \\ \hline
    \Tstrut $^{82}$Se & 34 & 48 & 4.718 & 2.3 & 0.1404 & 0.05786 & 0.06641 & 0.08169 & 0.09376 \\ \hline
    \Tstrut $^{79}$Br & 35 & 44 & 4.752 & 2.3 & 0.1455 & 0.05983 & 0.06901 & 0.07521 & 0.08675 \\ \hline
    \Tstrut $^{81}$Br & 35 & 46 & 4.747 & 2.3 & 0.14581 & 0.05997 & 0.06916 & 0.07882 & 0.09089 \\ \hline
    \Tstrut $^{80}$Kr & 36 & 44 & 4.802 & 2.3 & 0.14923 & 0.06117 & 0.07094 & 0.07476 & 0.08671 \\ \hline
    \Tstrut $^{82}$Kr & 36 & 46 & 4.794 & 2.3 & 0.14979 & 0.06143 & 0.07122 & 0.07849 & 0.091 \\ \hline
    \Tstrut $^{83}$Kr & 36 & 47 & 4.785 & 2.3 & 0.15045 & 0.06172 & 0.07154 & 0.08059 & 0.0934 \\ \hline
    \Tstrut $^{84}$Kr & 36 & 48 & 4.788 & 2.3 & 0.15021 & 0.06162 & 0.07142 & 0.08215 & 0.09523 \\ \hline
    \Tstrut $^{86}$Kr & 36 & 50 & 4.782 & 2.3 & 0.15068 & 0.06183 & 0.07165 & 0.08587 & 0.09951 \\ \hline
    \Tstrut $^{85}$Rb & 37 & 48 & 4.811 & 2.3 & 0.15587 & 0.06382 & 0.07431 & 0.08279 & 0.09641 \\ \hline
    \Tstrut $^{87}$Rb & 37 & 50 & 4.804 & 2.3 & 0.15642 & 0.06407 & 0.07458 & 0.08658 & 0.10079 \\ \hline
    \Tstrut $^{86}$Sr & 38 & 48 & 4.85 & 2.3 & 0.16011 & 0.06537 & 0.07652 & 0.08257 & 0.09665 \\ \hline
    \Tstrut $^{87}$Sr & 38 & 49 & 4.84 & 2.3 & 0.1609 & 0.06573 & 0.07691 & 0.08476 & 0.09917 \\ \hline
    \Tstrut $^{88}$Sr & 38 & 50 & 4.84 & 2.3 & 0.16094 & 0.06574 & 0.07692 & 0.08651 & 0.10121 \\ \hline
    \Tstrut $^{89}$Y\footnotemark[2] & 39 & 50 & 4.76 & 2.5092 & 0.16775 & 0.0686 & 0.0804 & 0.08795 & 0.10307 \\ \hline
    \Tstrut $^{90}$Zr & 40 & 50 & 4.901 & 2.3 & 0.16991 & 0.06904 & 0.0816 & 0.0863 & 0.10201 \\ \hline
    \Tstrut $^{91}$Zr & 40 & 51 & 4.929 & 2.3 & 0.1681 & 0.06822 & 0.08071 & 0.08698 & 0.10291 \\ \hline
    \Tstrut $^{92}$Zr & 40 & 52 & 4.958 & 2.3 & 0.1656 & 0.06709 & 0.07948 & 0.08722 & 0.10332 \\ \hline
    \Tstrut $^{94}$Zr & 40 & 54 & 4.995 & 2.3 & 0.16258 & 0.06572 & 0.07799 & 0.08873 & 0.10529 \\ \hline
    \Tstrut $^{96}$Zr & 40 & 56 & 5.021 & 2.3 & 0.16036 & 0.06473 & 0.0769 & 0.09062 & 0.10766 \\ \hline
    \Tstrut Isotope & $Z$ & $N$ & $c$ & $t$ & $D$ & $S(p)$ & $V(p)$ & $S(n)$ & $V(n)$ \\ \hline
    \Tstrut $^{93}$Nb & 41 & 52 & 4.985 & 2.3 & 0.17029 & 0.06882 & 0.08193 & 0.08729 & 0.10391 \\ \hline
    \Tstrut $^{92}$Mo & 42 & 50 & 4.975 & 2.3 & 0.17823 & 0.07201 & 0.086 & 0.08572 & 0.10238 \\ \hline
    \Tstrut $^{94}$Mo & 42 & 52 & 5.026 & 2.3 & 0.1736 & 0.06992 & 0.0837 & 0.08656 & 0.10363 \\ \hline
    \Tstrut $^{95}$Mo & 42 & 53 & 5.041 & 2.3 & 0.17231 & 0.06933 & 0.08307 & 0.08749 & 0.10482 \\ \hline
    \Tstrut $^{96}$Mo & 42 & 54 & 5.071 & 2.3 & 0.1696 & 0.06811 & 0.08172 & 0.08757 & 0.10507 \\ \hline
    \Tstrut $^{97}$Mo & 42 & 55 & 5.076 & 2.3 & 0.16919 & 0.06793 & 0.08152 & 0.08895 & 0.10675 \\ \hline
    \Tstrut $^{98}$Mo & 42 & 56 & 5.105 & 2.3 & 0.16656 & 0.06674 & 0.08022 & 0.08899 & 0.10696 \\ \hline
    \Tstrut $^{100}$Mo & 42 & 58 & 5.156 & 2.3 & 0.16209 & 0.06474 & 0.07802 & 0.08941 & 0.10774 \\ \hline
    \Tstrut $^{96}$Ru & 44 & 50 & 5.085 & 2.3 & 0.18189 & 0.07284 & 0.08811 & 0.08277 & 0.10012 \\ \hline
    \Tstrut $^{98}$Ru & 44 & 54 & 5.128 & 2.3 & 0.17771 & 0.07096 & 0.08603 & 0.08709 & 0.10559 \\ \hline
    \Tstrut $^{99}$Ru & 44 & 55 & 5.145 & 2.3 & 0.17613 & 0.07025 & 0.08525 & 0.08781 & 0.10656 \\ \hline
    \Tstrut $^{100}$Ru & 44 & 56 & 5.171 & 2.3 & 0.17362 & 0.06913 & 0.084 & 0.08798 & 0.10691 \\ \hline
    \Tstrut $^{101}$Ru & 44 & 57 & 5.183 & 2.3 & 0.17254 & 0.06864 & 0.08347 & 0.08892 & 0.10813 \\ \hline
    \Tstrut $^{102}$Ru & 44 & 58 & 5.211 & 2.3 & 0.16992 & 0.06747 & 0.08217 & 0.08894 & 0.10831 \\ \hline
    \Tstrut $^{104}$Ru & 44 & 60 & 5.252 & 2.3 & 0.16617 & 0.0658 & 0.08031 & 0.08972 & 0.10952 \\ \hline
    \Tstrut $^{103}$Rh\footnotemark[3] & 45 & 58 & 5.176 & 2.3 & 0.17963 & 0.07142 & 0.08713 & 0.09205 & 0.11231 \\ \hline
    \Tstrut $^{102}$Pd & 46 & 56 & 5.216 & 2.3 & 0.18221 & 0.07216 & 0.08857 & 0.08785 & 0.10782 \\ \hline
    \Tstrut $^{104}$Pd & 46 & 58 & 5.251 & 2.3 & 0.17869 & 0.07059 & 0.08682 & 0.089 & 0.10946 \\ \hline
    \Tstrut $^{105}$Pd & 46 & 59 & 5.261 & 2.3 & 0.17767 & 0.07013 & 0.08631 & 0.08995 & 0.1107 \\ \hline
    \Tstrut $^{106}$Pd & 46 & 60 & 5.285 & 2.3 & 0.17536 & 0.0691 & 0.08516 & 0.09014 & 0.11108 \\ \hline
    \Tstrut $^{108}$Pd & 46 & 62 & 5.318 & 2.3 & 0.17207 & 0.06764 & 0.08352 & 0.09117 & 0.11257 \\ \hline
    \Tstrut $^{110}$Pd & 46 & 64 & 5.349 & 2.3 & 0.16911 & 0.06633 & 0.08206 & 0.09228 & 0.11416 \\ \hline
    \Tstrut $^{107}$Ag & 47 & 60 & 5.301 & 2.3 & 0.17992 & 0.07073 & 0.08758 & 0.09029 & 0.1118 \\ \hline
    \Tstrut $^{109}$Ag & 47 & 62 & 5.331 & 2.3 & 0.1769 & 0.06938 & 0.08607 & 0.09153 & 0.11354 \\ \hline
    \Tstrut $^{106}$Cd & 48 & 58 & 5.288 & 2.3 & 0.18746 & 0.07367 & 0.0915 & 0.08902 & 0.11056 \\ \hline
    \Tstrut $^{110}$Cd & 48 & 62 & 5.344 & 2.3 & 0.18161 & 0.07107 & 0.08857 & 0.0918 & 0.11441 \\ \hline
    \Tstrut $^{111}$Cd & 48 & 63 & 5.351 & 2.3 & 0.18088 & 0.07075 & 0.08821 & 0.09285 & 0.11577 \\ \hline
    \Tstrut $^{112}$Cd & 48 & 64 & 5.371 & 2.3 & 0.1788 & 0.06983 & 0.08717 & 0.0931 & 0.11623 \\ \hline
    \Tstrut $^{113}$Cd & 48 & 65 & 5.378 & 2.3 & 0.17811 & 0.06952 & 0.08683 & 0.09414 & 0.11758 \\ \hline
    \Tstrut $^{114}$Cd & 48 & 66 & 5.395 & 2.3 & 0.17628 & 0.06871 & 0.08592 & 0.09447 & 0.11813 \\ \hline
    \Tstrut $^{116}$Cd & 48 & 68 & 5.416 & 2.3 & 0.17414 & 0.06776 & 0.08485 & 0.096 & 0.12021 \\ \hline
    \Tstrut $^{113}$In & 49 & 64 & 5.379 & 2.3 & 0.18388 & 0.07167 & 0.08986 & 0.0936 & 0.11737 \\ \hline
    \Tstrut $^{115}$In & 49 & 66 & 5.402 & 2.3 & 0.18146 & 0.0706 & 0.08865 & 0.09509 & 0.11941 \\ \hline
    \Tstrut $^{116}$Sn & 50 & 66 & 5.417 & 2.3 & 0.18565 & 0.07204 & 0.0909 & 0.09509 & 0.11999 \\ \hline
    \Tstrut $^{117}$Sn & 50 & 67 & 5.424 & 2.3 & 0.18491 & 0.07171 & 0.09054 & 0.09609 & 0.12132 \\ \hline
    \Tstrut $^{118}$Sn & 50 & 68 & 5.439 & 2.3 & 0.1833 & 0.071 & 0.08973 & 0.09655 & 0.12203 \\ \hline
    \Tstrut $^{119}$Sn & 50 & 69 & 5.443 & 2.3 & 0.18287 & 0.07081 & 0.08951 & 0.09771 & 0.12353 \\ \hline
    \Tstrut $^{120}$Sn & 50 & 70 & 5.459 & 2.3 & 0.18119 & 0.07007 & 0.08867 & 0.09809 & 0.12414 \\ \hline
    \Tstrut $^{122}$Sn & 50 & 72 & 5.476 & 2.3 & 0.17935 & 0.06925 & 0.08775 & 0.09973 & 0.12637 \\ \hline
    \Tstrut $^{124}$Sn & 50 & 74 & 5.491 & 2.3 & 0.1778 & 0.06857 & 0.08698 & 0.10149 & 0.12873 \\ \hline
    \Tstrut $^{121}$Sb & 51 & 70 & 5.496 & 2.3 & 0.18278 & 0.07036 & 0.08963 & 0.09657 & 0.12302 \\ \hline
    \Tstrut $^{123}$Sb & 51 & 72 & 5.508 & 2.3 & 0.18153 & 0.06981 & 0.089 & 0.09855 & 0.12565 \\ \hline
    \Tstrut $^{122}$Te & 52 & 70 & 5.537 & 2.3 & 0.18387 & 0.07042 & 0.09033 & 0.0948 & 0.1216 \\ \hline
    \Tstrut $^{124}$Te & 52 & 72 & 5.55 & 2.3 & 0.18238 & 0.06977 & 0.08958 & 0.0966 & 0.12404 \\ \hline
    \Tstrut $^{125}$Te & 52 & 73 & 5.553 & 2.3 & 0.18212 & 0.06966 & 0.08946 & 0.09779 & 0.12558 \\ \hline
    \Tstrut $^{126}$Te & 52 & 74 & 5.562 & 2.3 & 0.18112 & 0.06922 & 0.08895 & 0.0985 & 0.12659 \\ \hline
    \Tstrut $^{128}$Te & 52 & 76 & 5.573 & 2.3 & 0.17991 & 0.06869 & 0.08835 & 0.10039 & 0.12912 \\ \hline
    \Tstrut $^{130}$Te & 52 & 78 & 5.583 & 2.3 & 0.17878 & 0.06819 & 0.08778 & 0.10229 & 0.13167 \\ \hline
    \Tstrut $^{127}$I & 53 & 74 & 5.593 & 2.3 & 0.18299 & 0.06963 & 0.09005 & 0.09721 & 0.12573 \\ \hline
    \Tstrut Isotope & $Z$ & $N$ & $c$ & $t$ & $D$ & $S(p)$ & $V(p)$ & $S(n)$ & $V(n)$ \\ \hline
    \Tstrut $^{128}$Xe & 54 & 74 & 5.63 & 2.3 & 0.18405 & 0.06967 & 0.09074 & 0.09548 & 0.12435 \\ \hline
    \Tstrut $^{129}$Xe & 54 & 75 & 5.632 & 2.3 & 0.18392 & 0.06962 & 0.09068 & 0.09669 & 0.12594 \\ \hline
    \Tstrut $^{130}$Xe & 54 & 76 & 5.641 & 2.3 & 0.18282 & 0.06914 & 0.09013 & 0.09731 & 0.12684 \\ \hline
    \Tstrut $^{131}$Xe & 54 & 77 & 5.638 & 2.3 & 0.18314 & 0.06928 & 0.09028 & 0.09879 & 0.12874 \\ \hline
    \Tstrut $^{132}$Xe & 54 & 78 & 5.646 & 2.3 & 0.18228 & 0.0689 & 0.08985 & 0.09953 & 0.12979 \\ \hline
    \Tstrut $^{134}$Xe & 54 & 80 & 5.654 & 2.3 & 0.18139 & 0.06852 & 0.08941 & 0.1015 & 0.13245 \\ \hline
    \Tstrut $^{136}$Xe & 54 & 82 & 5.664 & 2.3 & 0.18025 & 0.06802 & 0.08883 & 0.10328 & 0.13489 \\ \hline
    \Tstrut $^{133}$Cs & 55 & 78 & 5.671 & 2.3 & 0.18447 & 0.06942 & 0.09109 & 0.09845 & 0.12918 \\ \hline
    \Tstrut $^{134}$Ba & 56 & 78 & 5.707 & 2.3 & 0.18538 & 0.0694 & 0.09171 & 0.09667 & 0.12774 \\ \hline
    \Tstrut $^{135}$Ba & 56 & 79 & 5.703 & 2.3 & 0.18576 & 0.06956 & 0.0919 & 0.09814 & 0.12964 \\ \hline
    \Tstrut $^{136}$Ba & 56 & 80 & 5.711 & 2.3 & 0.18485 & 0.06917 & 0.09144 & 0.09881 & 0.13063 \\ \hline
    \Tstrut $^{137}$Ba & 56 & 81 & 5.71 & 2.3 & 0.18494 & 0.06921 & 0.09149 & 0.10011 & 0.13233 \\ \hline
    \Tstrut $^{138}$Ba & 56 & 82 & 5.72 & 2.3 & 0.18382 & 0.06872 & 0.09092 & 0.10063 & 0.13313 \\ \hline
    \Tstrut $^{139}$La & 57 & 82 & 5.742 & 2.3 & 0.18613 & 0.06931 & 0.09225 & 0.09971 & 0.13271 \\ \hline
    \Tstrut $^{140}$Ce & 58 & 82 & 5.774 & 2.3 & 0.18723 & 0.06937 & 0.09297 & 0.09807 & 0.13144 \\ \hline
    \Tstrut $^{142}$Ce & 58 & 84 & 5.815 & 2.3 & 0.18229 & 0.06724 & 0.09048 & 0.09738 & 0.13103 \\ \hline
    \Tstrut $^{141}$Pr & 59 & 82 & 5.795 & 2.3 & 0.18947 & 0.06991 & 0.09426 & 0.09716 & 0.13101 \\ \hline \hline
    \Tstrut $^{203}$Tl & 81 & 122 & 6.602 & 2.3 & 0.16445 & 0.05061 & 0.08501 & 0.07622 & 0.12804 \\ \hline
    \Tstrut $^{205}$Tl & 81 & 124 & 6.617 & 2.3 & 0.16241 & 0.04979 & 0.08396 & 0.07622 & 0.12853 \\ \hline
    \Tstrut $^{204}$Pb & 82 & 122 & 6.617 & 2.3 & 0.16521 & 0.05052 & 0.08556 & 0.07516 & 0.12729 \\ \hline
    \Tstrut $^{206}$Pb & 82 & 124 & 6.631 & 2.3 & 0.16325 & 0.04974 & 0.08455 & 0.07521 & 0.12785 \\ \hline
    \Tstrut $^{207}$Pb & 82 & 125 & 6.637 & 2.3 & 0.16248 & 0.04943 & 0.08415 & 0.07536 & 0.12828 \\ \hline
    \Tstrut $^{208}$Pb & 82 & 126 & 6.637 & 2.3 & 0.16247 & 0.04943 & 0.08414 & 0.07595 & 0.12929 \\ \hline
    \Tstrut $^{209}$Bi & 83 & 126 & 6.687 & 2.3 & 0.15828 & 0.04738 & 0.08213 & 0.07193 & 0.12468 \\ \hline
\end{longtable}

\twocolumngrid

\vbox{}

\footnotetext[1]{Landolt-B\"{o}rnstein (only rms radius) \cite{Landolt-Bornstein2004}}
\footnotetext[2]{deVries et al.\cite{deVries1987}}
\footnotetext[3]{Angeli et al.\cite{Angeli2013}}

\clearpage

\onecolumngrid

\begin{longtable}{|c|c|c|c|c|c|c|c|c|c|c|c|c|c|}
\caption{\label{table:deformed}Table of overlap integrals for deformed nuclei}\\
\noalign{\vspace{1.5em}}
    \hline
    \Tstrut Isotope & $Z$ & $N$ & $c$ & $t_{eq}$ & $\beta$ & $\alpha$ & $k$ & $R_{k \alpha}$ & $D$ & $S(p)$ & $V(p)$ & $S(n)$ & $V(n)$ \\ \hline
    \Tstrut $^{142}$Nd & 60 & 82 & 5.825 & 2.3 & 0 & 0.124 & 2.243 & 6.284 & 0.1905 & 0.0699 & 0.095 & 0.0956 & 0.1298 \\ \hline
    \Tstrut $^{143}$Nd & 60 & 83 & 5.84 & 2.303 & 0.032 & 0.124 & 2.243 & 6.3 & 0.1885 & 0.0691 & 0.094 & 0.0956 & 0.13 \\ \hline
    \Tstrut $^{144}$Nd & 60 & 84 & 5.864 & 2.3 & 0 & 0.124 & 2.243 & 6.32 & 0.1857 & 0.0679 & 0.0925 & 0.095 & 0.1296 \\ \hline
    \Tstrut $^{145}$Nd & 60 & 85 & 5.881 & 2.341 & 0.109 & 0.124 & 2.243 & 6.351 & 0.1824 & 0.0665 & 0.0908 & 0.0942 & 0.1287 \\ \hline
    \Tstrut $^{146}$Nd & 60 & 86 & 5.902 & 2.383 & 0.154 & 0.124 & 2.243 & 6.386 & 0.1785 & 0.0649 & 0.0888 & 0.0931 & 0.1274 \\ \hline
    \Tstrut $^{148}$Nd & 60 & 88 & 5.944 & 2.46 & 0.213 & 0.124 & 2.243 & 6.454 & 0.1711 & 0.0618 & 0.085 & 0.0909 & 0.125 \\ \hline
    \Tstrut $^{150}$Nd & 60 & 90 & 5.882 & 2.758 & 0.371 & 0.138 & 2.346 & 6.52 & 0.1683 & 0.0609 & 0.0833 & 0.092 & 0.1258 \\ \hline
    \Tstrut $^{144}$Sm & 62 & 82 & 5.862 & 2.298 & 0 & 0.138 & 2.319 & 6.315 & 0.1951 & 0.0711 & 0.0976 & 0.094 & 0.1291 \\ \hline
    \Tstrut $^{147}$Sm & 62 & 85 & 5.945 & 2.275 & 0.14 & 0.138 & 2.319 & 6.382 & 0.1855 & 0.0669 & 0.0927 & 0.0917 & 0.1272 \\ \hline
    \Tstrut $^{148}$Sm & 62 & 86 & 6.008 & 2.176 & 0.156 & 0.135 & 2.295 & 6.405 & 0.1806 & 0.0646 & 0.0902 & 0.0898 & 0.1254 \\ \hline
    \Tstrut $^{149}$Sm & 62 & 87 & 5.981 & 2.29 & 0.183 & 0.136 & 2.304 & 6.421 & 0.1804 & 0.0648 & 0.0902 & 0.0909 & 0.1265 \\ \hline
    \Tstrut $^{150}$Sm & 62 & 88 & 5.856 & 2.607 & 0.222 & 0.125 & 2.221 & 6.431 & 0.1851 & 0.0671 & 0.0922 & 0.0957 & 0.1315 \\ \hline
    \Tstrut $^{152}$Sm & 62 & 90 & 5.869 & 2.741 & 0.309 & 0.133 & 2.288 & 6.499 & 0.1788 & 0.0645 & 0.0888 & 0.0944 & 0.13 \\ \hline
    \Tstrut $^{154}$Sm & 62 & 92 & 5.966 & 2.619 & 0.345 & 0.133 & 2.288 & 6.535 & 0.1717 & 0.0614 & 0.0854 & 0.0917 & 0.1275 \\ \hline
    \Tstrut $^{151}$Eu & 63 & 88 & 6.037 & 2.357 & 0.215 & 0.136 & 2.301 & 6.497 & 0.1757 & 0.0625 & 0.0879 & 0.0872 & 0.1228 \\ \hline
    \Tstrut $^{153}$Eu & 63 & 90 & 5.935 & 2.532 & 0.249 & 0.136 & 2.307 & 6.471 & 0.1825 & 0.0656 & 0.0913 & 0.0938 & 0.1305 \\ \hline
    \Tstrut $^{154}$Gd & 64 & 90 & 5.963 & 2.626 & 0.291 & 0.136 & 2.295 & 6.533 & 0.18 & 0.0641 & 0.0898 & 0.0908 & 0.1273 \\ \hline
    \Tstrut $^{155}$Gd & 64 & 91 & 6.02 & 2.403 & 0.249 & 0.14 & 2.332 & 6.498 & 0.1805 & 0.0642 & 0.0905 & 0.0912 & 0.1287 \\ \hline
    \Tstrut $^{156}$Gd & 64 & 92 & 6.009 & 2.658 & 0.341 & 0.137 & 2.311 & 6.589 & 0.1733 & 0.0613 & 0.0864 & 0.0889 & 0.1252 \\ \hline
    \Tstrut $^{157}$Gd & 64 & 93 & 6.01 & 2.475 & 0.271 & 0.135 & 2.292 & 6.516 & 0.1793 & 0.0638 & 0.0899 & 0.0927 & 0.1306 \\ \hline
    \Tstrut $^{158}$Gd & 64 & 94 & 6.023 & 2.678 & 0.35 & 0.137 & 2.31 & 6.608 & 0.1711 & 0.0604 & 0.0853 & 0.0894 & 0.1262 \\ \hline
    \Tstrut $^{160}$Gd & 64 & 96 & 6.061 & 2.645 & 0.358 & 0.132 & 2.268 & 6.629 & 0.1678 & 0.059 & 0.0837 & 0.089 & 0.1263 \\ \hline
    \Tstrut $^{159}$Tb\footnotemark[4] & 65 & 94 & 6.2 & 1.984 & 0.3 & 0.137\footnotemark[5] & 2.304\footnotemark[5] & 6.521 & 0.1738 & 0.0602 & 0.0873 & 0.0871 & 0.1263 \\ \hline
    \Tstrut $^{160}$Dy\footnotemark[3] & 66 & 94 & 6.092 & 2.712 & 0.338 & 0.139 & 2.306 & 6.682 & 0.1695 & 0.0591 & 0.0849 & 0.0847 & 0.1216 \\ \hline
    \Tstrut $^{161}$Dy & 66 & 95 & 6.038 & 2.564 & 0.271 & 0.139 & 2.306 & 6.574 & 0.181 & 0.0639 & 0.091 & 0.0919 & 0.1311 \\ \hline
    \Tstrut $^{162}$Dy & 66 & 96 & 6.042 & 2.725 & 0.347 & 0.139 & 2.307 & 6.642 & 0.1751 & 0.0615 & 0.0877 & 0.0899 & 0.1282 \\ \hline
    \Tstrut $^{163}$Dy & 66 & 97 & 6.044 & 2.587 & 0.283 & 0.139 & 2.31 & 6.589 & 0.1794 & 0.0632 & 0.0902 & 0.0929 & 0.1326 \\ \hline
    \Tstrut $^{164}$Dy & 66 & 98 & 6.058 & 2.753 & 0.358 & 0.138 & 2.308 & 6.67 & 0.1721 & 0.0603 & 0.0863 & 0.0899 & 0.1286 \\ \hline
    \Tstrut $^{165}$Ho\footnotemark[3] & 67 & 98 & 6.12 & 2.78 & 0.284 & 0.14 & 2.314 & 6.735 & 0.1676 & 0.0583 & 0.0843 & 0.0852 & 0.1233 \\ \hline
    \Tstrut $^{164}$Er\footnotemark[3] & 68 & 96 & 6.14 & 2.702 & 0.331 & 0.143 & 2.32 & 6.719 & 0.1713 & 0.0592 & 0.0862 & 0.0838 & 0.1222 \\ \hline
    \Tstrut $^{166}$Er & 68 & 98 & 6.098 & 2.742 & 0.351 & 0.143 & 2.331 & 6.698 & 0.1752 & 0.0609 & 0.0882 & 0.088 & 0.1275 \\ \hline
    \Tstrut $^{167}$Er & 68 & 99 & 6.108 & 2.622 & 0.297 & 0.142 & 2.319 & 6.659 & 0.178 & 0.0619 & 0.0898 & 0.0901 & 0.1308 \\ \hline
    \Tstrut $^{168}$Er & 68 & 100 & 6.121 & 2.764 & 0.359 & 0.141 & 2.316 & 6.728 & 0.1716 & 0.0595 & 0.0865 & 0.0876 & 0.1273 \\ \hline
    \Tstrut $^{170}$Er & 68 & 102 & 6.144 & 2.765 & 0.358 & 0.141 & 2.32 & 6.749 & 0.1689 & 0.0583 & 0.0851 & 0.0876 & 0.1278 \\ \hline
    \Tstrut $^{169}$Tm\footnotemark[6] & 69 & 100 & 6.126 & 2.567 & 0.298 & 0.141\footnotemark[5] & 2.295\footnotemark[5] & 6.652 & 0.1812 & 0.0628 & 0.0916 & 0.091 & 0.1328 \\ \hline
    \Tstrut $^{170}$Yb & 70 & 100 & 6.212 & 2.675 & 0.357 & 0.141\footnotemark[5] & 2.305\footnotemark[5] & 6.773 & 0.1705 & 0.058 & 0.0861 & 0.0831 & 0.1234 \\ \hline
    \Tstrut $^{171}$Yb & 70 & 101 & 6.214 & 2.535 & 0.299 & 0.141\footnotemark[5] & 2.305\footnotemark[5] & 6.721 & 0.1746 & 0.0596 & 0.0884 & 0.0859 & 0.1276 \\ \hline
    \Tstrut $^{172}$Yb & 70 & 102 & 6.227 & 2.657 & 0.349 & 0.141\footnotemark[5] & 2.305\footnotemark[5] & 6.779 & 0.1692 & 0.0575 & 0.0856 & 0.0858 & 0.1277 \\ \hline
    \Tstrut $^{173}$Yb & 70 & 103 & 6.234 & 2.539 & 0.3 & 0.141\footnotemark[5] & 2.305\footnotemark[5] & 6.741 & 0.172 & 0.0585 & 0.0871 & 0.086 & 0.1281 \\ \hline
    \Tstrut $^{174}$Yb & 70 & 104 & 6.2 & 2.72 & 0.336 & 0.141\footnotemark[5] & 2.305\footnotemark[5] & 6.779 & 0.1706 & 0.0581 & 0.0861 & 0.0866 & 0.1283 \\ \hline
    \Tstrut $^{176}$Yb & 70 & 106 & 6.271 & 2.61 & 0.328 & 0.141\footnotemark[5] & 2.305\footnotemark[5] & 6.801 & 0.1652 & 0.0557 & 0.0834 & 0.0847 & 0.1269 \\ \hline
    \Tstrut $^{175}$Lu\footnotemark[3] & 71 & 104 & 6.313 & 2.627 & 0.289 & 0.14\footnotemark[5] & 2.314\footnotemark[5] & 6.847 & 0.1629 & 0.0545 & 0.0826 & 0.0799 & 0.121 \\ \hline
    \Tstrut $^{176}$Lu\footnotemark[3] & 71 & 105 & 6.329 & 2.605 & 0.278 & 0.14\footnotemark[5] & 2.314\footnotemark[5] & 6.853 & 0.1616 & 0.0539 & 0.0819 & 0.0798 & 0.1212 \\ \hline
    \Tstrut $^{176}$Hf & 72 & 104 & 6.288 & 2.643 & 0.305 & 0.1388 & 2.274 & 6.828 & 0.1688 & 0.0567 & 0.0859 & 0.0817 & 0.1238 \\ \hline
    \Tstrut $^{177}$Hf & 72 & 105 & 6.297 & 2.575 & 0.277 & 0.1388 & 2.275 & 6.81 & 0.1697 & 0.0568 & 0.0862 & 0.0829 & 0.1258 \\ \hline
    \Tstrut $^{178}$Hf & 72 & 106 & 6.317 & 2.607 & 0.296 & 0.1388 & 2.275 & 6.84 & 0.1663 & 0.0555 & 0.0845 & 0.0816 & 0.1242 \\ \hline
    \Tstrut $^{179}$Hf & 72 & 107 & 6.348 & 2.485 & 0.267 & 0.1388 & 2.276 & 6.824 & 0.1659 & 0.0551 & 0.0843 & 0.0819 & 0.1253 \\ \hline
\Tstrut Isotope & $Z$ & $N$ & $c$ & $t_{eq}$ & $\beta$ & $\alpha$ & $k$ & $R_{k \alpha}$ & $D$ & $S(p)$ & $V(p)$ & $S(n)$ & $V(n)$ \\ \hline
    \Tstrut $^{180}$Hf & 72 & 108 & 6.318 & 2.57 & 0.286 & 0.1457 & 2.331 & 6.827 & 0.1673 & 0.0558 & 0.085 & 0.0837 & 0.1274 \\ \hline
    \Tstrut $^{181}$Ta & 73 & 108 & 6.367 & 2.446 & 0.255 & 0.1462 & 2.327 & 6.826 & 0.1678 & 0.0554 & 0.0855 & 0.082 & 0.1265 \\ \hline
    \Tstrut $^{182}$W & 74 & 108 & 6.334 & 2.594 & 0.273 & 0.1464 & 2.32 & 6.848 & 0.171 & 0.0567 & 0.0873 & 0.0826 & 0.1271 \\ \hline
    \Tstrut $^{183}$W\footnotemark[3] & 74 & 109 & 6.343 & 2.536 & 0.243 & 0.147 & 2.33 & 6.837 & 0.1715 & 0.0567 & 0.0875 & 0.0835 & 0.1288 \\ \hline
    \Tstrut $^{184}$W & 74 & 110 & 6.36 & 2.569 & 0.26 & 0.1483 & 2.358 & 6.865 & 0.1684 & 0.0556 & 0.086 & 0.0824 & 0.1275 \\ \hline
    \Tstrut $^{186}$W & 74 & 112 & 6.384 & 2.559 & 0.254 & 0.1465 & 2.324 & 6.882 & 0.1656 & 0.0544 & 0.0845 & 0.0822 & 0.1277 \\ \hline
    \Tstrut $^{185}$Re\footnotemark[3] & 75 & 110 & 6.36 & 2.498 & 0.221 & 0.1458\footnotemark[5] & 2.311\footnotemark[5] & 6.837 & 0.1737 & 0.0571 & 0.0888 & 0.0838 & 0.1302 \\ \hline
    \Tstrut $^{187}$Re\footnotemark[3] & 75 & 112 & 6.374 & 2.498 & 0.221 & 0.1458\footnotemark[5] & 2.311\footnotemark[5] & 6.85 & 0.1718 & 0.0564 & 0.0878 & 0.0842 & 0.1311 \\ \hline
    \Tstrut $^{186}$Os\footnotemark[7] & 76 & 110 & 6.414 & 2.572 & 0.261 & 0.1445 & 2.28 & 6.911 & 0.1676 & 0.0545 & 0.0859 & 0.0787 & 0.124 \\ \hline
    \Tstrut $^{187}$Os\footnotemark[3] & 76 & 111 & 6.416 & 2.48 & 0.209 & 0.1445 & 2.285 & 6.88 & 0.17 & 0.0552 & 0.087 & 0.0806 & 0.1271 \\ \hline
    \Tstrut $^{188}$Os & 76 & 112 & 6.427 & 2.561 & 0.243 & 0.1445 & 2.289 & 6.92 & 0.1663 & 0.0539 & 0.0852 & 0.0793 & 0.1253 \\ \hline
    \Tstrut $^{189}$Os\footnotemark[3] & 76 & 113 & 6.435 & 2.463 & 0.198 & 0.1445 & 2.29 & 6.892 & 0.168 & 0.0543 & 0.086 & 0.0808 & 0.1279 \\ \hline
    \Tstrut $^{190}$Os & 76 & 114 & 6.456 & 2.452 & 0.194 & 0.1445 & 2.292 & 6.907 & 0.1656 & 0.0534 & 0.0849 & 0.08 & 0.1271 \\ \hline
    \Tstrut $^{192}$Os & 76 & 116 & 6.483 & 2.37 & 0.169 & 0.1445 & 2.293 & 6.904 & 0.1642 & 0.0526 & 0.0841 & 0.0803 & 0.1283 \\ \hline
    \Tstrut $^{191}$Ir\footnotemark[3] & 77 & 114 & 6.45 & 2.413 & 0.164 & 0.1455\footnotemark[5] & 2.293\footnotemark[5] & 6.888 & 0.1704 & 0.0548 & 0.0874 & 0.0812 & 0.1294 \\ \hline
    \Tstrut $^{193}$Ir\footnotemark[3] & 77 & 116 & 6.472 & 2.384 & 0.141 & 0.1455\footnotemark[5] & 2.293\footnotemark[5] & 6.898 & 0.1683 & 0.0539 & 0.0864 & 0.0812 & 0.1301 \\ \hline
    \Tstrut $^{194}$Pt & 78 & 116 & 6.537 & 2.387 & -0.145 & 0.1465 & 2.298 & 6.959 & 0.1625 & 0.0513 & 0.0836 & 0.0761 & 0.1241 \\ \hline
    \Tstrut $^{195}$Pt & 78 & 117 & 6.541 & 2.373 & 0.13 & 0.1465 & 2.298 & 6.958 & 0.1624 & 0.0511 & 0.0835 & 0.0767 & 0.1252 \\ \hline
    \Tstrut $^{196}$Pt & 78 & 118 & 6.547 & 2.369 & -0.129 & 0.1464 & 2.298 & 6.963 & 0.1616 & 0.0508 & 0.0832 & 0.0768 & 0.1256 \\ \hline
    \Tstrut $^{198}$Pt & 78 & 120 & 6.56 & 2.349 & -0.108 & 0.1463 & 2.299 & 6.968 & 0.1605 & 0.0503 & 0.0825 & 0.0774 & 0.1269 \\ \hline
    \Tstrut $^{197}$Au & 79 & 118 & 6.554 & 2.365 & -0.125 & 0.1477 & 2.3 & 6.967 & 0.1637 & 0.0512 & 0.0843 & 0.0765 & 0.1259 \\ \hline
    \Tstrut $^{198}$Hg & 80 & 118 & 6.569 & 2.357 & -0.116 & 0.1488 & 2.303 & 6.978 & 0.1647 & 0.0513 & 0.0851 & 0.0755 & 0.1252 \\ \hline
    \Tstrut $^{199}$Hg & 80 & 119 & 6.571 & 2.356 & -0.115 & 0.1488 & 2.303 & 6.979 & 0.1645 & 0.0512 & 0.0849 & 0.0761 & 0.1262 \\ \hline
    \Tstrut $^{200}$Hg & 80 & 120 & 6.582 & 2.341 & -0.098 & 0.1487 & 2.303 & 6.985 & 0.1633 & 0.0507 & 0.0843 & 0.0759 & 0.1263 \\ \hline
    \Tstrut $^{201}$Hg\footnotemark[3] & 80 & 121 & 6.583 & 2.33 & -0.084 & 0.1486 & 2.303 & 6.982 & 0.1635 & 0.0507 & 0.0844 & 0.0767 & 0.1276 \\ \hline
    \Tstrut $^{202}$Hg & 80 & 122 & 6.596 & 2.327 & -0.079 & 0.1486 & 2.303 & 6.992 & 0.1619 & 0.0501 & 0.0836 & 0.0763 & 0.1274 \\ \hline
    \Tstrut $^{204}$Hg & 80 & 124 & 6.611 & 2.319 & -0.066 & 0.1485 & 2.303 & 7.002 & 0.1601 & 0.0493 & 0.0826 & 0.0764 & 0.128 \\ \hline \hline
    \Tstrut $^{232}$Th\footnotemark[8] & 90 & 142 & 6.945 & 2.581 & 0.263 & 0.165\footnotemark[5] & 2.344\footnotemark[5] & 7.387 & 0.1344 & 0.0363 & 0.0707 & 0.0572 & 0.1113 \\ \hline
    \Tstrut $^{238}$U\footnotemark[9] & 92 & 146 & 7.011 & 2.627 & 0.288 & 0.1626\footnotemark[5] & 2.35\footnotemark[5] & 7.467 & 0.129 & 0.0337 & 0.0681 & 0.0534 & 0.108 \\ \hline
\end{longtable}

\twocolumngrid

\vbox{}
\footnotetext[1]{Landolt-B\"{o}rnstein (only rms radius) \cite{Landolt-Bornstein2004}}
\footnotetext[2]{deVries et al.\cite{deVries1987}}
\footnotetext[3]{Angeli et al.\cite{Angeli2013}}
\footnotetext[4]{deWit et al. \cite{Dewit1966}}
\footnotetext[5]{$\alpha$ and $k$ are interpolated using neighboring elements}
\footnotetext[6]{Landolt-B\"ornstein \cite{Landolt-Bornstein2004}}
\footnotetext[7]{Hoehn et al. \cite{Hoehn1981}}
\footnotetext[8]{Zumbro et al. (1986) \cite{Zumbro1986}}
\footnotetext[9]{Zumbro et al. (1984) \cite{Zumbro1984}}

\clearpage

\section{\label{sec:errors_apB}Comparison of overlap  integral determinations}

In three instances,  $_{27}^{13}$Al, $_{\:\:58}^{142}$Ce and $_{\:\:82}^{208}$Pb, we have been able to compare  our calculation with earlier work and alternative experimental 
inputs~\cite{Kitano2002, Heeck2022, Zhang2022, Qweak2021}. 

In calculating the CLFV matrix elements, Kitano {\it et al.} primarily used neutron distributions equal to proton distributions. In a few cases, they also compared the $S(n)$ and $V(n)$ values determined using direct measurements from pionic atoms and proton scattering. They also assigned an uncertainty in the pionic atom case based on the uncertainty of the $C_n$ parameter. A recent measurement of the neutron radius of aluminum~\cite{Qweak2021} provides a third example.

\begin{figure}[!ht]
    \includegraphics[width=0.5\textwidth]{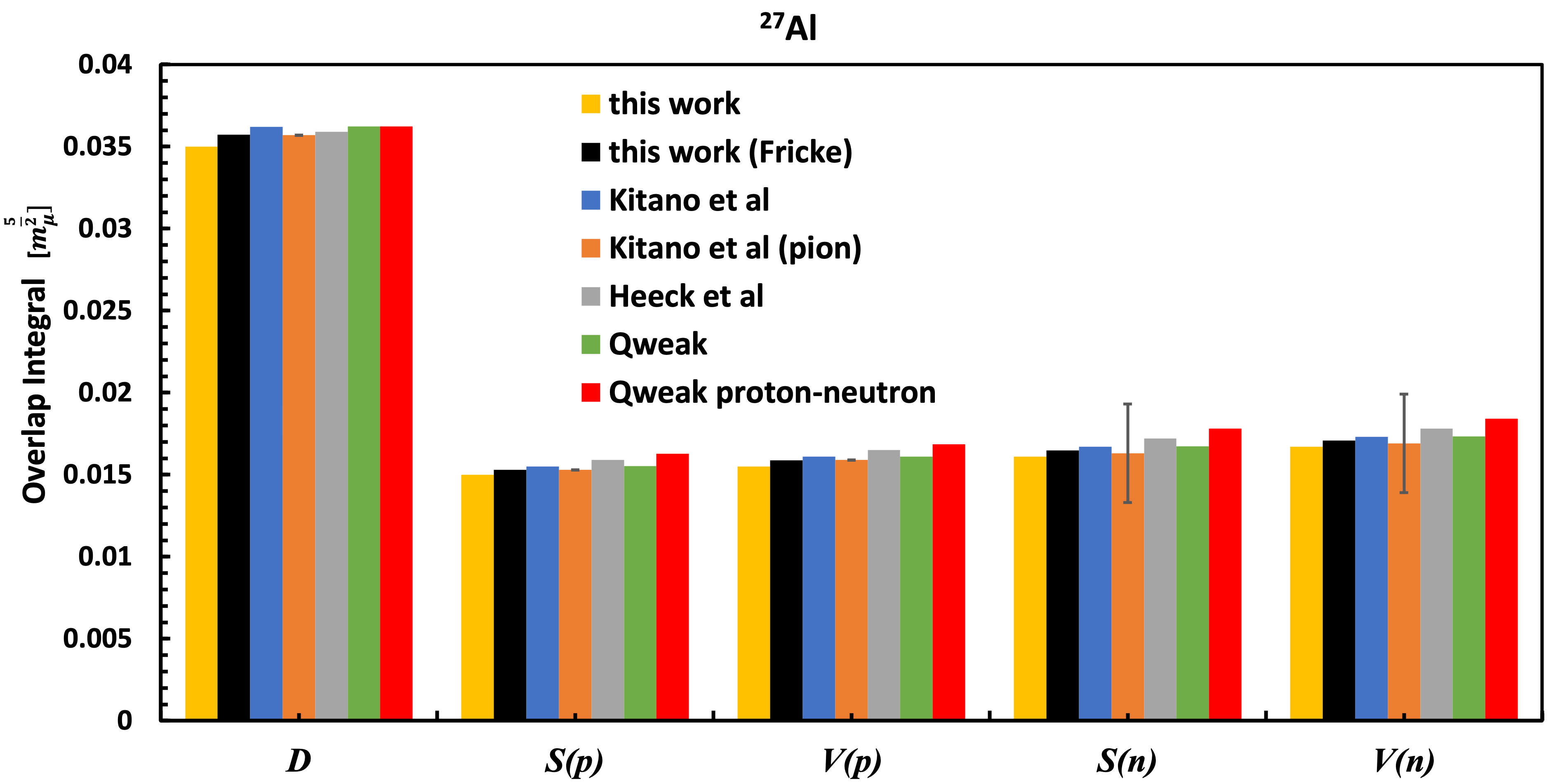}
    \caption{Comparison of CLFV overlap integral values for aluminum using $N/Z$ scaling for the neutron overlap integral with experimental measurements and the Zhang {\it et al.} model. }
    \label{fig:Al_comp}
\end{figure}

\begin{figure}[!ht]
    \includegraphics[width=3.5in]{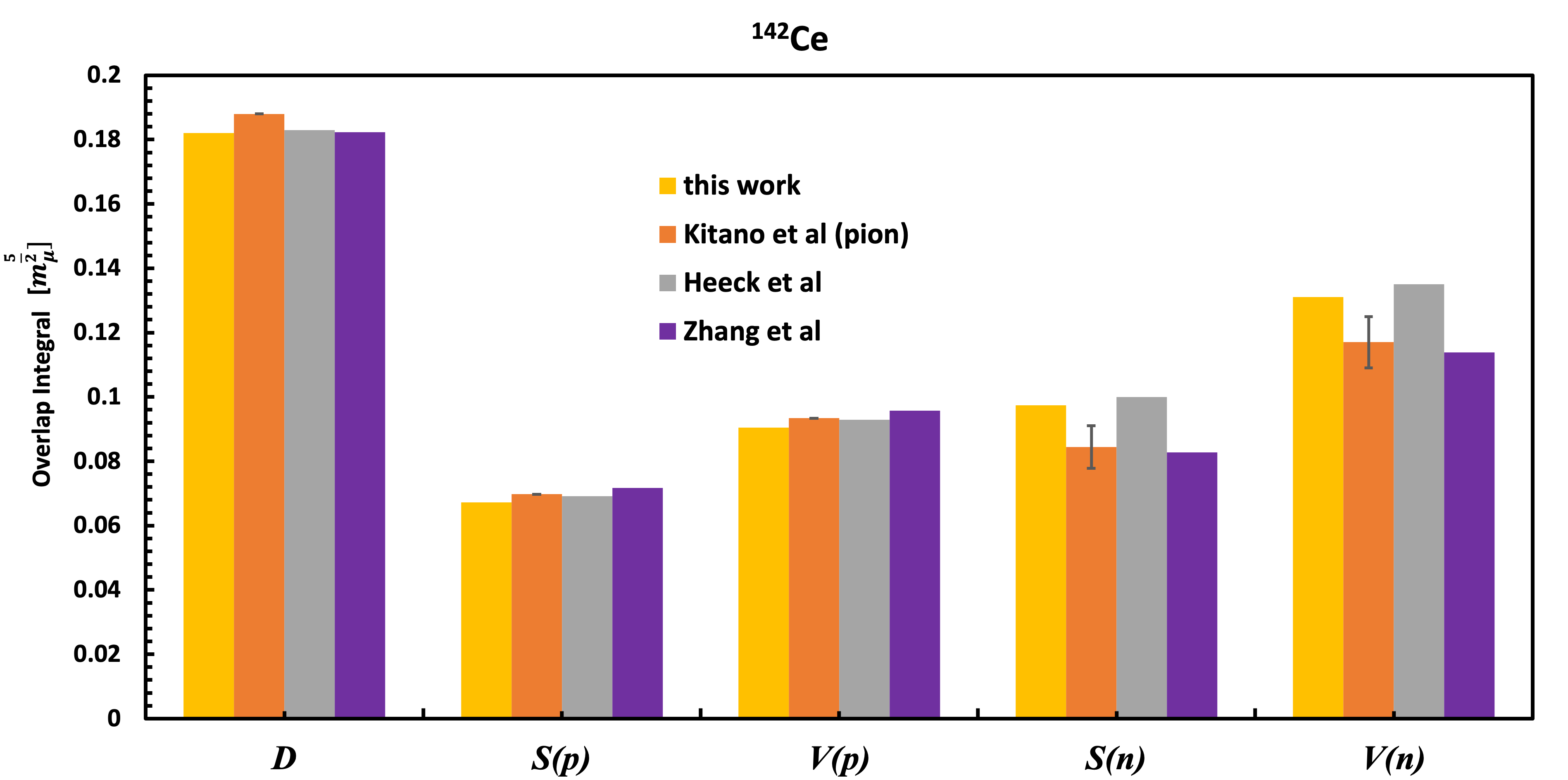}
    \caption{Comparison of CLFV overlap integral values for cerium using $N/Z$ scaling for the neutron overlap integral with experimental measurements and the Zhang {\it et al.} model. }
    \label{fig:Ce_comp}
\bigskip
    \includegraphics[width=3.5in]{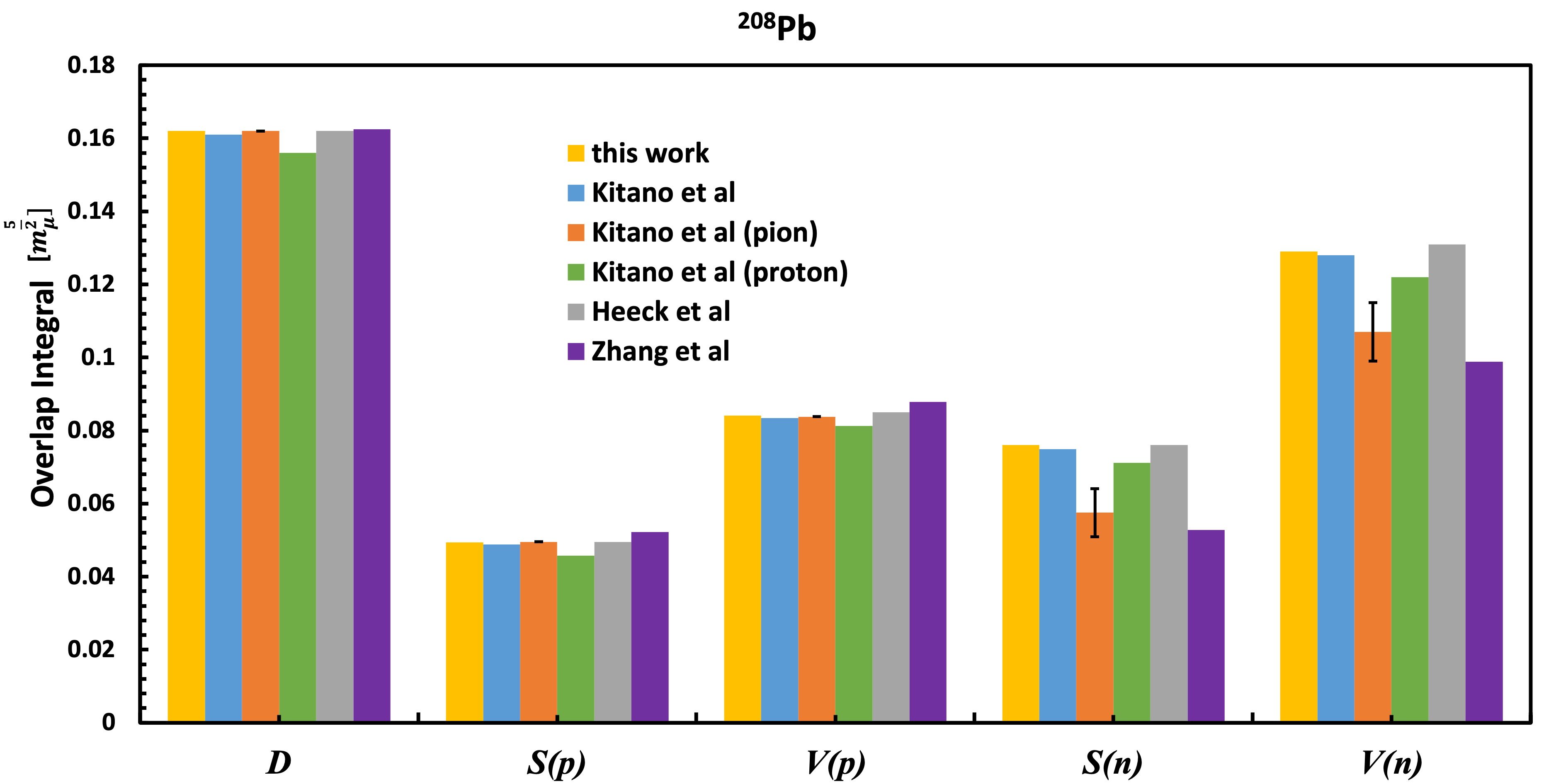}
    \caption{Comparison of CLFV overlap integral values for lead using $N/Z$ scaling for the neutron overlap integral with experimental measurements and the Zhang {\it et al.} model. }
    \label{fig:Pb_comp}
\end{figure}

Figures~\ref{fig:Al_comp} -- \ref{fig:Pb_comp} show comparisons of the overlap integrals for $_{27}^{13}$Al, $_{\:\:58}^{142}$Ce and $_{\:\:82}^{208}$Pb from the different calculations. Ce and Pb are spherical nuclei and Al has a quadrupole moment $\beta = -0.392$. The overlap integrals for these nuclei are not surprisingly in excellent agreement. The uncertainties assigned to the pionic atom neutron distribution by Kitano {\it et al.} are shown. It is notable that the $S(n)$ and $V(n)$ values Kitano {\it et al.} that use the pionic atom measurements are in agreement with the values obtained using the Zhang {\it et al.} model for the neutron distribution.  For aluminum, we use the recent measurement of the neutron radius via parity-violating electron scattering~\cite{Qweak2021}\cite{Horowitz1999}. Since $^{27}$Al is an odd-even nucleus, there is no Zhang {\it et al.} result. We, therefore, use the neutron skin calculation of~\cite{Sammarruca2022}.

The results for aluminum are not particularly sensitive to the neutron model used, as there is only one more neutron than proton in the nucleus. For cerium and lead, the use of experimental measurements of the neutron distribution rather than proton distribution leads to better agreement with the Zhang {\it et al.} model results, providing support for our approach.

\clearpage

\bibliographystyle{apsrev4-2}
\bibliography{main}
\end{document}